\documentclass[12pt]{article}
\usepackage{amsmath}
\usepackage{amsfonts}
\usepackage{epsfig}
\renewcommand{\phi}{\varphi}

\newcommand{\beq}{\begin{equation}}
\newcommand{\eeq}{\end{equation}}

\newcommand{\pd}{\partial}

\newcommand{\bphi}{\bar{\phi}}

\newcommand{\bg}{\bar{g}}
\newcommand{\bdel}{\bar{\delta}}
\newcommand{\balpha}{\bar{\alpha}}

\newcommand{\ep}{\epsilon}
\newcommand{\lra}{\leftrightarrow}
\newcommand{\ra}{\rho_{10}}
\newcommand{\rb}{\rho_{20}}
\newcommand{\rc}{\rho_{30}}

\title{\bf On the reggeon model with the pomeron and odderon:
singularities with non-zero masses.}
\author{M.A. Braun$^1$, E.M. Kuzminskii$^{2}$, M.I. Vyazovsky$^1$\\
$^1${\small Dept. of High Energy Physics, Saint-Petersburg State University,
St. Petersburg, Russia}\\
$^2${\small Theoretical Physics Division,
Petersburg Nuclear Physics Institute, Gatchina, Russia}}

\begin{document}
\maketitle
{\bf Abstract}

\noindent
\noindent
The Regge-Gribov model of the pomeron and odderon in the non-trivial
transverse space is studied by the renormalization group technique in the single loop approximation.
The pomeron and odderon are taken to have different bare intercepts and slopes.
The behaviour when the intercepts move from below to their
critical values compatible with the Froissart limitation is studied.
The singularities in the form of non-trivial branch points indicating a phase transition
are found in the vicinity of five fixed points found in the previous publication.
Since new phases violate the projectile-target symmetry the model is found non-physical
for the bare intercepts above their critical value.

\section{Introduction}
In the kinematic region
where the energy is much greater than transferred momenta (''the Regge
kinematics'')  strong interactions can be phenomenologically
described by the exchange of reggeons, which correspond to poles in the
complex angular momentum plane. In this framework the high-energy asymptotic
is governed by the exchange of pomerons with a pole intercept  close to unity $\alpha_P(0)=1$.
Further development leads to interaction between  pomerons conveniently
described by the theory introduced by V.N.Gribov with the triple pomeron vertex and an
imaginary coupling constant. Much attention was given to the study of this
theory in the past \cite{gribov,migdal,migdal1}. This theory was also long ago  applied to the study of
the $pA$ interaction at high energies in ~\cite{schwimmer}, where the sum of all
fan diagrams  was found (similar to the later treatment in the QCD framework,
which lead to the well-known Balitski-Kovchegov equations ~\cite{bal,kov1,kov2}).

Being essentially simpler than the QCD approach, the reggeon theory is, however, still
a full-fledged quantum field theory and does not allow to find constructively scattering
amplitudes. To achieve this goal a  simpler model in the zero-dimensional transverse world
(''toy'' model) was considered and studied in some detail
\cite{amati1,amati2,jengo,amati3,rossi,braun1,bondarenko,braun3,braun2}.
This model   is essentially equivalent to the standard quantum
mechanics and can be studied by its well developed methods.  The important messages which
followed from these studies were that 1) the quantum effects, that is the
loops, change cardinally the high-energy behaviour of the amplitudes
and so their neglect is at most a very crude approximation
and 2) passage through the intercept $\alpha_P(0)=1$ goes
smoothly, without phase transition, so that the theory preserves its physical sense
for the supercritical pomeron with $\alpha_P(0)>1$.

The second of these important findings has been, however, found  wrong in the physical
case of two transverse dimensions. Using the renormalization group (RG) technique in
\cite{abarb2} it was concluded that at $\alpha_P(0)=1$ a second order phase transition occurs.
New phases, which arise at $\alpha_P(0)>1$, cannot be considered  physical, since they violate
the fundamental symmetry target-projectile. So the net result was that the model
could not accomodate the supercritical pomeron with $\alpha_P(0)>1$ altogether.

In the QCD, apart from the pomeron with the
positive $C$-parity and signature, a compound state of three reggeized gluons with
the negative $C$-parity and signature, the odderon, appears. Actually, it was proposed before the QCD era
on general grounds in ~\cite{nicol}. Since then its  possible experimental
manifestations has been widely discussed
~\cite{odderon1,odderon2,odderon3} with  conculsions containing a large dose of uncertainty up to now, which may be
explained both by the difficulties in the experimental settings and  the elusive properties of the odderon itself.
On the theoretical level the QCD  odderon was discussed in many papers \cite{jw1,jw2,blv,hiim}.
Its intercept was found to lie in the vicinity of unity and as was noted
in \cite{hiim} that the  odderon may in a certain sense constitute an imaginary
part of the full $S$-matrix with charge parities $C=\pm 1$ exchanges whose real part
is the pomeron. So the coupling constants for the odderon interactions are
probably the same as for the pomeron interactions.

In the reggeon field approach  we introduced the odderon into the
zero-dimensional Regge model to study the influence of the odderon on the properties
of the model  ~\cite{bkv}. Our numerical results have shown that this influence is minimal.
No phase transition occurs as both intercepts cross unity and the cross-sections
continue to slowly diminish at high energies whether intercepts are smaller or greater than unity.

In the realistic two-dimensional
transverse world within the functional RG approach the reggeon theory with the odderon was considered in
~\cite{bartels, vacca}
 where two of the five real fixed points were found and
the corresponding general structure of the pomeron-odderon interaction was analysed.
More detailed study within the standard perturbative RG framework was made in our paper
~\cite{bkv2023} for the massless reggeons. We found five real fixed points (and several complex ones).
In the single loop approximation they turned out to be only partially attractive and
the study of evolution showed that the coupling constants either go to the three of the
five fixed points or go away indicating the loss of precision.
However, since the masses were initially taken zero (corresponding to the original
intercept exactly equal to unity) the problem of the transition above this value was left open.

In this note  using the RG approach we study the model with  odderons in two transverse dimensions
with masses different from zero both for the pomeron and odderon.
As in ~\cite{bkv2023} we limit ourselves with the lowest non-trivial (single loop) approximation.
Our aim is to see what happens when either of the two masses vanishes (that is the original intercept goes to unity).
It turns out that at zero masses observables have branch points, continuation beyond which leads to appearance of
two complex conjugated singularitues thus indicating a phase transition and developing a non-zero vacuum expectation value
of the reggeon fields. Since the odderon field cannot have a non-zero expectation value, the situation will be the same one
as happens without odderon in ~\cite{abarb2}. The new  phase will violate
the projectile-target symmetry exactly as without odderons and has to be discarded.
So the presence of the odderon will not improve the model and prohibit intercepts to become greater than unity.

We also study evolution of the scattering amplitudes at high energies taking into account coupling of the system to participants.
We find that the dominant contributions come from the exchange of  single full propagators of the pomeron or odderon, with all interaction
taken into account in them.
The corresponding cross-sections behave as $(\ln s)^{1/6}$ for the pomeron exchange and $(\ln s)^{1/12}$ for the odderon exchange.

\section{Model. Renormalization and evolution}
Our model describes two  fields $\phi_{1,2}$ for the pomeron $\phi_1$ and odderon $\phi_2$ acting in the D-dimensional transverse space
with the Lagrangian
\[
{\cal L}=\sum_{i=1}^2\Big(\bphi_{i0}\pd_y\phi_{i0}-\mu_{i0}\bphi_{i0}\phi_{i0}+\alpha'_{i0}\nabla\bphi_{i0}\nabla\phi_{i0}\Big)\]\beq
+\frac{i}{2}\Big(\lambda_{10}\bphi_{10}(\phi_{10}+\bphi_{10})\phi_{10}+2\lambda_{20}(\bphi_{20}\phi_{20}(\bphi_{10}+\phi_{10}))
+\lambda_{30}(\bphi_{20}^2\phi_{10}-\phi_{20}^2\bphi_{10})\Big).
\label{eq1}
\eeq
It contains two different bare "masses" $\mu_{10}$ and $\mu_{20}$ and slope parameters $\alpha'_{i0}$ for the pomeron and odderon.
The masses are defined as the intercepts minus unity. In the free theory with $\lambda_l=0$ one has
$\alpha_i(0)=1+\mu_{i0}$, $i=1,2$.
With $\mu<0$ simple perturbation approach is
effective and for $\mu>0$ the theory is badly defined, does not admit direct summation of perturbation series and needs analytic continuation.
As found in ~\cite{abarb2} for the theory without odderon such continuation is prohibited on physical grounds. We postpone investigation of whether
presence of the odderon can improve the situation  for future studies. The number of dimensions relevant for the application of the
RG technique is $D=4-\ep$ with $\ep\to 0$. Physically, of course, $D=2$.
This  theory  is invariant under transformation
\beq \phi_1(y,x)\lra \bphi_1(-y,x),\ \ \phi_2(y,x)\lra i\bphi_2(-y,x),
\label{invj4}
\eeq
which reflects the symmetry between the projectile and target.
It has to be supplemented by the external coupling to participants in the form
\beq
{\cal L}_{ext}=i\rho_p(x)\phi_1(Y/2,x)+i\rho_t(x)\bphi_1(-Y/2,x)+\rho_p^{(O)}(x)\phi_2(Y/2,x)+i\rho_t^{(O)}(x)\bphi_2(-Y/2,x),
\label{lext}
\eeq
with the amplitude ${\cal A}$  given by
\beq
{\cal A}_{pt}(Y)=-i\Big<{\rm T}\Big\{e^{\int d^2x{\cal L}_{ext}} S_{int}\Big\}\Big> ,
\label{ampli}
\eeq
where $S_{int}$ is the standard $S$ matrix in the interaction representation.
A rather peculiar form for the interaction of the odderon to the participants arises due to specific canonical
transformation of the odderon fields made to simplify its interactions.

We introduce Green functions without external legs,
that is multiplied by the inverse  propagator for each leg, which
 are  characterized by numbers $m_1,m_2$  and $n_1,n_2$ of reggeons before and
after interaction
\[\Gamma^{n_1,n_2,m_1,m_2}(E, k,\alpha'_{j0},\lambda_{l0}),\ \ j=1,2,\ \ l=1,2,3.\]
In fact $\Gamma$ may depend on several energies and momenta of initial and final reggeons.
To economize on notations we denote the whole set of them as $E$ and $k$ meaning
\[E=\{E_1,E_2,...\},\ \ k=\{{k}_1,{k}_2,...\},\ \ k^2=\{ {k}_i {k}_j \},\ \ i=1,2,...\ \ j=1,2,...\ .\]
Also in the following the superscript $\{n_1,n_2,m_1,m_2\}$ will be suppressed except the special
cases when the concrete numbers $n_i$ and $m_i$ are important.
Our special interest will be in the two inverse propagators
\[\Gamma_1=\Gamma^{1,0,1,0}\ \  {\rm and}\ \ \Gamma_2=\Gamma^{0,1,0,1}.\]

Following \cite{abarb2} we introduce the lowest
 eigenvalue $M_i(\mu_{10},\mu_{20}))$ of the Hamiltonian for the pomeron and odderon
as the point where the inverse propagators $\Gamma_i(E,k^2)$ vanish
\beq
\Gamma_i(E,k^2)|_{E=M_i(\mu_{10},\mu_{20}),k=0}=0,\ \ i=1,2.
\label{eq4}
\eeq
Singularities at  $E=M$ are not supposed to be  isolated poles in the full propagator $G^{(2)}(E, k^2)$ but rather branch-points
resulting from the pole and all Regge cuts.

We assume that similar to the case without odderon ~\cite{abarb2}  $M_1(\mu_{10},\mu_{20})$, initially positive, diminishes as $\mu_{10}$
grows up to its maximal value
$\mu_{10c}$ at which $M_1$ reaches its critical value $M_{1c}=0$ compatible with the Froissart bound, as occurs in the perturbative approach.
This suggests introducing instead of  $\mu_{10}$  a variable $\delta_{10}$
\[\delta_{10}=\mu_{10c}-\mu_{10},\]
which is initially non-negative and vanishes when $\mu_{10}$ and $M_1$ attain their critical values at fixed $\mu_{20}$.
Note that in the free theory with $\lambda_l=0$ we evidently have
\[\Gamma_{1,\lambda=0}(E,k^2)=E-\alpha_{10}'k^2+\mu_{10}\]
so that
$\Gamma_{1,\lambda=0,k=0}=E+\mu_{10}$
and
$M_{1,\lambda=0}(\mu_{10},\mu_{20})=-\mu_{10}$.
It becomes equal to zero at $\mu_{10}=0$. As a result, in the free theory
$\mu_{10c}=0$, which means that in the presence of interaction $\mu_{10c}$ is of the second order
in $\lambda$ and corresponds to mass renormalization.

Similarly for $M_2(\mu_{10},\mu_{20})$ it is convenient to determine the value $\mu_{20c}$ at which $M_2$ attains its minimal value $M_2=0$
at fixed $\mu_{10}$ and define a non-negative variable $\delta_{20}$ as the difference
\[\delta_{20}=\mu_{20c}-\mu_{20}.\]
Values of both mass renormalization constants $\mu_{10c}$ and $\mu_{20c}$  will be  determined from (\ref{eq4}).
Note that the chosen scheme of renormalization with the subtraction
of unrenormalized critical mass allows one to avoid the mass mixing
and so simplifies the RG equations.

Renormalized quantities are introduced in the standard manner:
\[\phi_i=Z_i^{-1/2}\phi_{i0},\ \ i=1,2,\]
\[\alpha'_i=U_{i}^{-1}Z_i\alpha'_{i0},\ \ i=1,2,\]
\[\delta_i=T_i^{-1}Z_i\delta_{i0},\ \ i=1,2,\]
\[\lambda_1=W_{1}^{-1}Z_{1}^{3/2}\lambda_{10},\ \ \lambda_{2,3}=W_{2,3}^{-1}Z_1^{1/2}Z_2\lambda_{20,30},\]
where we have denoted $W$ the  standard vertex normalization constant
and $U$ and $T$ new renormalization constants for the slopes and masses.

The generalized vertices transform as
\[\Gamma^{R,n_1n_2,m_1,m_2}(E,k,\lambda_i,\alpha'_i,\delta_i, E_N)=Z_1^{(n_1+m_1)/2}Z_2^{(n_2+m_2)/2}
\Gamma^{n_1,m_1,n_2,m_2}(E,k,\lambda_{i0},\alpha'_{i0},\delta_{i0}),\]
where $E_N$ is the renormalization energy point.

Constants $Z$, $U$ $T$ and $W$ are determined by the renormalization conditions imposed on renormalized quantities,
which we borrow from ~\cite{abarb2,abarb1} suitably generalized to include the odderon:

\[\frac{\pd}{\pd E}\Gamma_i^R(E,k^2,\lambda,\alpha', E_N)\Big|_{E=-E_N,k^2=\delta_j=0}=1,\ \ i,j=1,2;\]\[
\frac{\pd}{\pd k^2}\Gamma_i^R(E,k^2,\lambda,\alpha', E_N)\Big|_{E=-E_N,k^2=\delta_j=0}=-\alpha'_i,\ \ i,j=1,2;\]\beq
\frac{\pd}{\pd \delta_i}\Gamma_i^R(E,k^2,\lambda,\alpha', E_N)\Big|_{E=-E_N,k^2=\delta_j=0}=-1\ \ i,j=1,2;
\label{renormc}
\eeq
\[\Gamma^{R,1,0,2,0}(E_i,k_i,\lambda_i,\alpha'_j,\delta_j, E_N)\Big|_{E_1=2E_2=2E_3=-E_N, k_j=\delta_j=0}=i\lambda_1 (2\pi)^{-(D+1)/2},
\ \ i=1,2,3,\ \ j=1,2;\]
\[\Gamma^{R,0,1,1,1}(E_i,k_i,\lambda_i,\alpha'_j,\delta_j, E_N)\Big|_{E_1=2E_2=2E_3=-E_N, k_j=\delta_j=0}=i\lambda_2 (2\pi)^{-(D+1)/2},
\ \ i=1,2,3,\ \ j=1,2;\]
\[\Gamma^{R,1,0,0,2}(E_i,k_i,\lambda_i,\alpha'_j, \delta_j,E_N)\Big|_{E_1=2E_2=2E_3=-E_N, k_j=\delta_j=0}=i\lambda_3 (2\pi)^{-(D+1)/2},
\ \ i=1,2,3,\ \ j=1,2\]
and we recall that the mass renormalization parameters $\mu_{i0c}$  are determined by the condition (\ref{eq4}):
\[\Gamma_i(E,k^2,\lambda_{i0},\alpha'_{10},\alpha'_{20},\delta_{10},\delta_{20})\Big|_{E=k^2=\delta_{i0}=0}=0,\ \ i=1,2.\]
Note that due to our definitions of $\delta_{10}$ and $\delta_{20}$ function $\Gamma_1$ vanishes at $\delta_{10}=0$
and $\Gamma_2$ at $\delta_{20}=0$.

We introduce new dimensionless coupling constants: unrenormalized  $u_0$ and renormalized~$u$
\[ g_{40}\equiv u_0=\frac{\alpha'_{20}}{\alpha'_{10}},\ \ g_4\equiv u=\frac{\alpha'_{2}}{\alpha'_1}.\]
The relation between them is determined as
\[ u=u_0\frac{Z_2U_1}{Z_1U_2}\equiv Z_4u_0.\]

With these normalizations the renormalization constants $Z$, $U$ $T$ and $W$  depend only on the dimensionless coupling constants
\beq
 g_i=\frac{\lambda_i}{(8\pi\alpha'_1)^{D/4}E_N^{(4-D)/4}},\ \ i=1,2,3\ \ {\rm and}\ \  g_4\equiv u.
 \label{gi}
\eeq

The RG equations  are standardly obtained from the condition that the unrenormalized $\Gamma$ do not depend on $E_N$.
So differentiating $\Gamma^R$ with respect to $E_N$
we get
\beq
\Big(E_N\frac{\pd}{\pd E_N}+\sum_{i=1}^4\beta_i(g)\frac{\pd}{\pd g_i}+
\sum_{i=1}^{2}\kappa_i(g)\delta_i\frac{\pd}{\pd \delta_i}+\tau_1(g)\alpha'_1\frac{\pd}{\pd\alpha'_1}
-\sum_{i=1}^2\frac{1}{2}(n_i+m_i)\gamma_i(g)\Big)\Gamma^{R}=0,
\label{eq44}
\eeq
where
\[
\beta_i(g)=E_N\frac{\pd g_i}{\pd E_N},\ \ i=1,...,4,\]
\[\gamma_i(g)=E_N\frac{\pd \ln Z_i}{\pd E_N},\ \ i=1,2,\]
\[\tau_i(g)=E_N\frac{\pd}{\pd E_N}\ln\Big(U_{i}^{-1}Z_i\Big),\ \ i=1,2,\]
\[\kappa_i(g)=E_N\frac{\pd}{\pd E_N}\ln\Big(T_{i}^{-1}Z_i\Big),\ \ i=1,2\]
and the derivatives are taken at $\lambda_{i0}$, $ u_0$, $\delta_{i0}$  and $\alpha'_{10}$  fixed.
For brevity  we denote in the following
\[\gamma(g)=\sum_{i=1}^2\frac{1}{2}(n_i+m_i)\gamma_i(g).\]
From the dimensional analysis we get
\[[\phi_i]=[\bphi_i]=k^{D/2},\ \ [\alpha'_i]=Ek^{-2},\ \ [\delta_i]=E,\ \ i=1,2,\]
\[
\Big[\Gamma^{R}\Big]=Ek^{D-(n+m)D/2},\ \ n=n_1+n_2,\ \ m=m_1+m_2.\]
This allows to write
\beq
\Gamma^{R}(E,k,g,\alpha'_1,\delta_{1,2},E_N)=E_N\Big(\frac{E_N}{\alpha'_1}\Big)^{(2-n-m)D/4}
\Phi\Big(\frac{E}{E_N},\frac{\alpha'_1}{E_N}k^2, \frac{\delta_{1,2}}{E_N},g\Big).
\eeq.

Using the scale transformation
\[E\to \frac{E}{\xi},\ \ k\to k\]
we find from the scale invariance
\beq
\Gamma^{R}(E,k^2,g,\alpha'_1,\delta_{1,2},E_N)=
\xi\Gamma^{R}\Big(\frac{E}{\xi},k^2,g,\frac{\alpha'_1}{\xi},\frac{\delta_{1,2}}{\xi},\frac{E_N}{\xi}\Big).
\label{gamxi}
\eeq

Our next procedure meets with the difficulty of having only one scale invariance  with two different $\delta_i$, $i=1,2$.
So we may take two different ways to scale only one of $\delta_{1,2}$ or both simultaneously.
In the following we adopt the first alternative and either scale $\delta_1$ leaving $\delta_2$ as an evolving variable
or scale $\delta_2$ with $\delta_1$ evolving.
In this way our procedure becomes a direct generalization of the pure pomeron case in~\cite{abarb2}.

So begin with substituting $\delta_1$ by $\xi\delta_1$ in (\ref{gamxi}). We get
\beq
\Gamma^{R}(E,k^2,g,\alpha',\xi\delta_1,\delta_2,E_N)=
\xi\Gamma^{R}\Big(\frac{E}{\xi},k^2,g,\frac{\alpha'}{\xi},\delta_1,\frac{\delta_2}{\xi},\frac{E_N}{\xi}\Big).
\label{eq51}
\eeq
Differentiation  by $\xi$
gives
\[
\xi\frac{\pd}{\pd \xi}\Gamma^{R}(E,k^2,g,\alpha',\xi\delta_1,\delta_2,E_N)=\]\beq
\Big(1-\alpha'_1\frac{\pd}{\pd\alpha'_1}
-\delta_2\frac{\pd}{\pd \delta_2}-E_N\frac{\pd}{\pd E_N}-E\frac{\pd}{\pd E}\Big)
\Gamma^{R}(E,k^2,g,\alpha',\xi\delta_1,\delta_2,E_N).
\label{eq52}
\eeq
Here
\[E\frac{\pd}{\pd E}=\sum_iE_i\frac{\pd}{\pd E_i},\ \ i=1,2,...\ .\]
From (\ref{eq44}) we find
\beq
\Big(\sum_{i=1}^4\beta_i(g)\frac{\pd}{\pd g_i}+\tau_1(g)\alpha'_1\frac{\pd}{\pd\alpha'_1}+\kappa_1(g)\delta_1\frac{\pd}{\pd\delta_1}
+\kappa_2(g)\delta_2\frac{\pd}{\pd\delta_2}
-\gamma(g)\Big)\Gamma^{R}=-E_N\frac{\pd}{\pd E_N}\Gamma^R.
\label{eq44a}
\eeq
This relation does not change if $\delta_1 \to\xi\delta_1$, so we can
put the left-hand side instead of $-E_N\pd/\pd E_N$ into (\ref{eq52}) to obtain
\[
\xi\frac{\pd}{\pd \xi}\Gamma^R(E,k^2,g,\alpha',\xi\delta_1,\delta_2,E_N)=
\Big(1-\alpha_1'\frac{\pd}{\pd\alpha'_1}-\delta_2\frac{\pd}{\pd \delta_2}-E\frac{\pd}{\pd E}+\sum_{i=1}^4\beta_i(g)\frac{\pd}{\pd g_i}\]\[
+\tau(g)\alpha'_1\frac{\pd}{\pd\alpha'_1}+\kappa_1(g)\delta\frac{\pd}{\pd\delta_1}
+\kappa_2(g)\delta_2\frac{\pd}{\pd\delta_2}
-\gamma(g)
\Big)\Gamma^R(E,k^2,g,\alpha',\xi\delta_1,\delta_2,E_N).\]

Next we note that acting on $\Gamma^R(E,k^2,g,\alpha'_1,\xi\delta_1,\delta_2,E_N)$
\[\delta_1\frac{\pd}{\pd\delta_1}=\xi\frac{\pd}{\pd \xi}=\xi\delta_1\frac{\pd}{\pd(\xi\delta_1)},\]
so that
\[
\xi\frac{\pd}{\pd \xi}\Gamma^R(E,k^2,g,\alpha',\xi\delta_1,\delta_2,E_N)=
\Big(1-\alpha'_1\frac{\pd}{\pd\alpha'_1}-\delta_2\frac{\pd}{\pd \delta_2}-E\frac{\pd}{\pd E}\]\[
+\sum_{i=1}^4\beta_i(g)\frac{\pd}{\pd g_i}+\tau_1(g)\alpha'\frac{\pd}{\pd\alpha'_1}+\kappa_1(g)\xi\frac{\pd}{\pd\xi}
+\kappa_2(g)\delta_2\frac{\pd}{\pd\delta_2}
-\gamma(g)
\Big)
\Gamma^R(E,k^2,g,\alpha',\xi\delta_1,\delta_2,E_N).\]
Transferring all terms to the left we find
\[
\Big([1-\kappa_1(g)]\xi\frac{\pd}{\pd\xi}-\sum_{i=1}^4\beta_i(g)\frac{\pd}{\pd g_i}+[1-\tau_1(g)]\alpha'_1\frac{\pd}{\pd\alpha'_1}\]
\[+[1-\kappa_2(g)]\delta_2\frac{\pd}{\pd\delta_2}
+E\frac{\pd}{\pd E}-[1-\gamma(g)]\Big)\Gamma^R(E,k^2,g,\alpha',\xi\delta_1,\delta_2,E_N)=0.\]
The solution of this equation is standard.
We put
\[t=\ln\xi.\]
Then
\[
\Gamma^R(E,k^2,g,\alpha'_1,\xi\delta_1,\delta_2,E_N)=
\Gamma^R\Big(\bar{E}(-t),k^2,\bar{g}(-t),\bar{\alpha}'_1(-t),\delta_1,\bar{\delta}_2(-t),E_N\Big)\]
\beq
\times
\exp\Big\{\int_{-t}^0dt'\frac{1-\gamma(\bar{g}(t'))}{1-\kappa_1(\bar{g}(t'))}\Big\},
\label{eq54}
\eeq
where
\[
\frac{d \bar{g}_i(t)}{dt}=-\frac{\beta_i(\bar{g}(t))}{1-\kappa_1(\bar{g}(t))},\]
\[\frac{d\ln\bar{\alpha}'_1(t)}{dt}=\frac{1-\tau_1(\bar{g}(t))}{1-\kappa_1(\bar{g}(t))},\]
\[\frac{d\ln\bar{\delta}_2(t)}{dt}=\frac{1-\kappa_2(\bar{g}(t))}{1-\kappa_1(\bar{g}(t))},\]
\beq
\frac{d\ln\bar{E}(t)}{dt}=\frac{1}{1-\kappa_1(\bar{g}(t))}
\label{eq55}
\eeq
with the initial conditions
\[\bar{g}_i(0)=g_i,\ \ \bar{\alpha}_1'(0)=\alpha'_1,\ \ \bar{\delta}_2(0)=\delta_2,\ \ \bar{E}(0)=E.\]


\section{Self-masses, anomalous dimensions and $\beta$-functions}
\subsection{Self-masses and renormalization constants $Z$, $U$ and $T$}
In this study, as mentioned, we restrict ourselves with the lowest order (single loop) approximation.

The unrenormalized inverse full propagators have the form
\beq
\Gamma_j(E,k^2)=E-\delta_{j0}-\alpha'_{j0}k^2+\mu_{j0c}-\Sigma_j(E,k^2),\ \ j=1,2,
\label{m0}
\eeq
where $\Sigma_j$ are the non-trivial self-masses. In the lowest approximation they are
graphically shown in Fig.~1.
\begin{figure}
\begin{center}
\epsfig{file=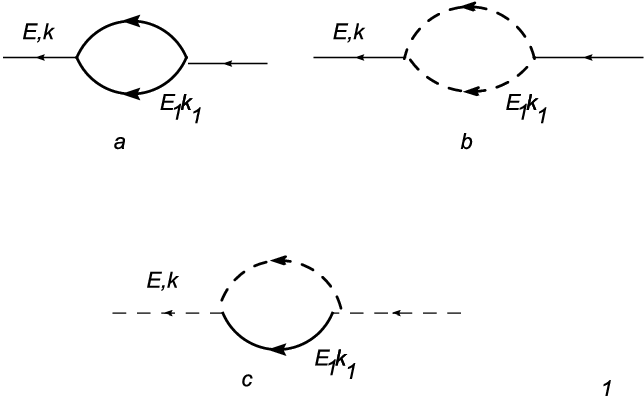, width=14 cm}
\caption{Self masses for $\Gamma_1$ ($a+b$) and $\Gamma_2$ ($c$). Pomerons and odderons are shown by solid and dashed lines respectively.}
\end{center}
\label{fig1}
\end{figure}
The unrenormalized self-masses are expressed via the unrenormalized parameters $\lambda_{l0}$ and $\alpha'_{i0}$
However, in the lowest order there is no difference between the renormalized and unrenormalized parameters and we use
the former ones.

 Condition (\ref{eq4}) has the form
\[\Gamma_j(E,k^2)\Big|_{E=k^2=\delta_{j0}=0}=\mu_{j0c}-\Sigma_j(E,k^2)_{E=k^2=\delta_{j0}=0}=0,\ \ j=1,2,\]
which determines $\mu_{10c}(\mu_{20})$ and $\mu_{20c}(\mu_{10})$ as
\beq
\mu_{j0c}=\Sigma_j(E,k^2)_{E=k=\delta_j=0},\ \ j=1,2.
\label{mu1c}
\eeq
So $\mu_{j0c}$ is fully determined as a subtraction term in $\Sigma_j$.
We denote
\[\Sigma_j(E,k^2)-\Sigma_j(E,k^2)_{E=k=\delta_j=0}=\tilde{\Sigma}_j(E,k^2),\ \ j=1,2.\]

The renormalized functions $\Gamma_j^R$ are defined as
\beq
\Gamma_j^R=Z_j\Gamma_j=Z_j(E-\alpha'_{j0}k^2-\delta_{j0}-\tilde{\Sigma}_j)
=Z_jE-U_j\alpha'_{j}k^2-T_j\delta_j-\tilde{\Sigma}_j(E,k^2),
\label{gamren}
\eeq
where we put $Z_j=1$ in front of $\Sigma_j$ having in mind the lowest non-trivial order.
The new constants $T_j$ are determined by the renormalization condition (\ref{renormc}) which gives
\beq
T_j-1=-\frac{\pd}{\pd \delta_j}\tilde{\Sigma}_j(E=-E_N,k=\delta_1=\delta_2=0),\ \ j=1,2.
\label{ti}
\eeq
We can rewrite (\ref{gamren}) as
\[\Gamma_j^R=E-\alpha'_j k^2-\delta_j+\Big((Z_j-1)E-(U_{j}-1)\alpha'_jk^2-(T_j-1)\delta_j-\tilde{\Sigma}_j(E,k^2)\Big).\]
The quantity in the bracket is the renormalized mass (with the opposite sign)
\beq
\Sigma_j^R=\tilde{\Sigma}_j-(Z_j-1)E+(U_j-1)\alpha'_j k^2+(T_j-1)\delta_j,
\label{sigr}
\eeq
so that
\beq
\Gamma_j^R=E-\alpha'_jk^2-\delta_j-\Sigma^R_j(E,k^2),\ \ j=1,2.\label{gamr}\eeq

We start from $\Gamma_1(E,k^2)$.

Consider the piece $\Sigma_1^a$. Explicitly
\[\Sigma_1^a=+\frac{1}{2}\lambda_{1}^2\int\frac{dE_1d^Dk_1}{2\pi i(2\pi)^D}\frac{1}{[E_1-\delta_1-\alpha'_{1}k_1^2+i0]
[E-E_1-\delta_1-\alpha'_{1}(k-k_1)^2+i0]}
\]
\[=-\frac{1}{2}\lambda_{1}^2\int\frac{d^Dk_1}{(2\pi)^D}\frac{1}{E-2\delta_1-\alpha_{1}'[k_1^2-(k_1-k)^2]}
=\frac{1}{2}\frac{\lambda_{1}^2}{2\alpha'_{1}}\int\frac{d^Dk_1}{(2\pi)^D}\frac{1}{k_1^2+a^2},\]
where
\[a^2=\frac{1}{4}k^2-\frac{E-2\delta_1}{2\alpha'_{1}}.\]
Calculating the integral we find the old result without mass with $E\to E-2\delta_1$

\beq
\Sigma_1^a=\frac{1}{2}g_1^2E_N^{2-D/2}\Gamma(1-D/2)\Big(\frac{1}{2}\alpha'_{1}k^2-E+2\delta_1\Big)^{D/2-1}.
\label{sig1a}
\eeq
At $E=k=\delta_1=0$ this expression vanishes provided $D/2-1>0$. So $\mu_{10c}^a=0$ and $\tilde{\Sigma}_1^a=\Sigma_1^a$.

The second piece $\Sigma_1^b$ is given by a similar expression with $\lambda_1\to\lambda_3$, $\alpha'_1\to\alpha'_2$,
$\delta_1\to\delta_2$ and opposite sign
\beq
\Sigma_1^b=-\frac{1}{2}\frac{g_3^2E_N^{2-D/2}}{u^{D/2}}\Gamma(1-D/2)\Big(\frac{1}{2}\alpha'_2k^2-E+2\delta_2\Big)^{D/2-1}.
\label{sig1b}
\eeq
From this we find
\[\mu_{10c}^b=-\frac{1}{2}\frac{g_3^2E_N^{2-D/2}}{u^{D/2}}\Gamma(1-D/2)(2\delta_2)^{D/2-1}\]
and
\[\tilde{\Sigma}_1^b=-\frac{1}{2}\frac{g_3^2E_N^{2-D/2}}{u^{D/2}}\Gamma(1-D/2)
\Big((\alpha'_2k^2/2-E+2\delta_2)^{D/2-1}-(2\delta_2)^{D/2-1}\Big).\]

To build the renormalized $\Sigma_1^R$ we need $Z_1,U_1$ and $T_1$.
As to the first two constants, they are determined at $\delta_1=\delta_2=0$ and so are the same as with
zero masses
\[Z_1^a-1=\frac{1}{2}g_1^2\Gamma(2-D/2),\ \ U_1^a-1=\frac{1}{4}g_1^2\Gamma(2-D/2),\]
\[Z_1^b-1=-\frac{1}{2}\frac{g_3^2}{u^{D/2}}\Gamma(2-D/2),\ \ U_1^b-1=-\frac{1}{4}\frac{g_3^2}{u^{D/2}}u\Gamma(2-D/2).\]

Now constant $T_1=1+(T^a_1-1)+(T^b_1-1)$. Differentiating in $\delta_1$ we find
\[T_1^a-1=-g_1^2E_N^{2-D/2}\Gamma(1-D/2)(D/2-1)E_N^{D/2-2}=g_1^2\Gamma(2-D/2),\]
\[T_1^b-1=0.\]

So the first piece of the renormalized self-mass $\Sigma^{Ra}$ will be is given as a sum
\[\Sigma_1^{Ra}=\frac{1}{2}g_1^2E_N^{2-D/2}\Gamma(1-D/2)\Big(\frac{1}{2}\alpha'_{1}k^2-E+2\delta_1\Big)^{D/2-1}
+\frac{1}{2}g_1^2\Gamma(2-D/2)(-E+\alpha'_1k^2/2+2\delta_1).\]
Denoting
\beq
\sigma_1=\frac{1}{2}\alpha'_{1}k^2+2\delta_1-E
\label{sms1}
\eeq
we obtain
\beq
\Sigma_1^{Ra}=\frac{1}{2}g_1^2\Gamma(2-D/2)\sigma_1\Big[\frac{(\sigma_1/E_N)^{D/2-2}}{1-D/2}+1\Big].
\label{sigr1a}
\eeq

Now the second piece $\Sigma_1^b$. It will be given by the sum
\[\Sigma_1^{Rb}=-\frac{1}{2}\frac{g_3^2E_N^{2-D/2}}{u^{D/2}}\Gamma(1-D/2)
\Big((\alpha'_2k^2/2-E+2\delta_2)^{D/2-1}-(2\delta_2)^{D/2-1}\Big)\]\[
-\frac{1}{2}\frac{g_3^2}{u^{D/2}}\Gamma(2-D/2)(-E+u\alpha'_1k^2/2).\]
We introduce
\beq\sigma_3=\frac{1}{2}\alpha'_2k^2-E+2\delta_2
\label{sms3}
\eeq
and rewrite the second term as
\[-\frac{1}{2}\frac{g_3^2}{u^{D/2}}\Gamma(2-D/2)(\sigma_3-2\delta_2).\]
This allows to find $\Sigma_1^{Rb}$ as a sum of two terms, each finite at $D=4$:
\beq
\Sigma^{Rb}=-\frac{1}{2}\frac{g_3^2}{u^{D/2}}\Gamma(2-D/2)\Big[\sigma_3\Big(\frac{(\sigma_3/E_N)^{D/2-2}}{1-D/2}+1\Big)
-2\delta_2\Big(\frac{(2\delta_2/E_N)^{D/2-2}}{1-D/2}+1\Big)\Big].
\label{sigr1b}
\eeq

We pass to the self-mass in $\Gamma_2$ shown in Fig.~1$c$.
We have
\[\Sigma_2=+\lambda_2^2\int\frac{dE_1d^Dk_1}{2\pi i(2\pi)^D}\frac{1}{[E_1-\alpha'_{1}k_1^2-\delta_1+i0][E-E_1-\alpha'_{2}(k-k_1)^2-\delta_2+i0]}
\]\[=-\lambda_2^2\int\frac{d^Dk_1}{(2\pi)^D}\frac{1}{E-\delta_1-\delta_2-\alpha_{1}'k_1^2-\alpha'_2(k_1-k)^2}.\]
In the denominator we find
\[\alpha'_1(1+u)(k_2^2+a^2),\]
where now
\[k_2=k_1-k\frac{u}{1+u},\ \ a^2=\frac{uk^2}{(1+u)^2}-\frac{E-\delta_1-\delta_2}{\alpha'_1(1+u)}.\]
As a result, comparing to the massless case we find
\beq
\Sigma_2=\frac{g_2^2E_N^{2-D/2}}{[(1+u)/2]^{D/2}}\Gamma(1-D/2)\Big(\alpha'_1k^2\frac{u}{1+u}-E+\delta_1+\delta_2\Big)^{D/2-1}.
\label{sig2}
\eeq
From (\ref{sig2}) we get
\[\mu_{20c}=\frac{g_2^2E_N^{2-D/2}}{[(1+u)/2]^{D/2}}\Gamma(1-D/2)\delta_1^{D/2-1},\]
so that after the mass renormalization and defining
\beq
\sigma_2=\alpha'_1k^2\frac{u}{1+u}-E+\delta_1+\delta_2
\label{sms2}
\eeq
we obtain
\beq
\tilde{\Sigma}_2=\frac{g_2^2E_N^{2-D/2}}{[(1+u)/2]^{D/2}}\Gamma(1-D/2)\Big(\sigma_2^{D/2-1}-\delta_1^{D/2-1}\Big).
\label{sig2a}
\eeq

Constants $Z_2$ and $U_2$ are calculated with $\delta_i=0$ and are the same as with zero masses:
\[Z_2-1=\frac{g_2^2}{[(1+u)/2]^{D/2}}\Gamma(2-D/2),\ \ U_2-1=\frac{g_2^2}{[(1+u)/2]^{D/2}}\Gamma(2-D/2)\frac{1}{1+u}.\]
We have
\[T_2-1=-\frac{\pd}{\pd \delta_2}\tilde{\Sigma}_2(E=-E_N,k=\delta_1=\delta_2=0)=\frac{g_2^2}{[(1+u)/2]^{D/2}}\Gamma(2-D/2).\]
So the renormalized $\Sigma_2$ will be given by the sum
\[\Sigma^R_2=\frac{g_2^2E_N^{2-D/2}}{[(1+u)/2]^{D/2}}\Gamma(1-D/2)\Big(\sigma_2^{D/2-1}-\delta_1^{D/2-1}\Big)\]\[
+\frac{g_2^2}{[(1+u)/2]^{D/2}}\Gamma(2-D/2)(-E+\alpha'_2k^2/(1+u)+\delta_2).\]
We rewrite the second term as
\[\frac{g_2^2}{[(1+u)/2]^{D/2}}\Gamma(2-D/2)(\sigma_2-\delta_1)\]
and present $\Sigma_2^R$ in the form
\beq
\frac{g_2^2}{[(1+u)/2]^{D/2}}\Gamma(2-D/2)\Big[\sigma_2\Big(\frac{(\sigma_2/E_N)^{D/2-2}}{1-D/2}+1\Big)
-\delta_1\Big(\frac{(\delta_1/E_N)^{D/2-2}}{1-D/2}+1\Big)\Big].
\label{sigR2}
\eeq
Both terms are finite at $D=4$. So renormalization procedure turns out to be correct.

\subsection{Anomalous dimensions, $\beta$-functions and fixed points}

To find the anomalous dimensions we have to differentiate the renormalization constants over $E_N$.
In the lowest order we have for all renormalized constants
\[\frac{\pd}{\pd E_N}\ln Z=\frac{\pd}{\pd E_N}\ln (1+Z-1)=\frac{\pd}{\pd E_N}(Z-1).\]
All renormalized constants depend on $E_N$
via constants $g_i$, $i=1,2,3$ and $g_4=u$, which in the lowest order are equal to the unrenormalized $g_{i0}$.
For $i=1,2,3$ one has
\[ E_N\frac{\pd}{\pd E_N}g_i^2=E_N\frac{\pd}{\pd E_N}\frac{\lambda_{i0}^2}{(8\pi\alpha'_{10})^{D/2}}E_N^{D/2-2}=
(D/2-2)\frac{\lambda_{i0}^2}{(8\pi\alpha'_{10})^{D/2}}E_N^{D/2-2}=(D/2-2)g_i^2,\]
whereas $u=\alpha'_2/\alpha'_1$ does not depend on $E_N$ in this order.
So to find the anomalous dimensions we have only to multiply the renormalization constants by $(D/2-2)$.
Each of them contains $\Gamma(2-D/2)$. So we shall have a product
\[(D/2-2)\Gamma(2-D/2)=-\Gamma(3-D/2).\]

Since $Z$ and $U$ are calculated at $\delta_1=\delta_2=0$, the anomalous dimensions $\gamma$ and
$\tau$ are the same as in the massless case from which we borrow
\beq
\gamma_1^a=-\frac{1}{2}g_1^2\Gamma(3-D/2),\label{gamma1a}\eeq
\beq
\gamma_1^b=\frac{1}{2}\frac{g_3^2}{u^{D/2}}\Gamma(3-D/2),\label{gamma1b}\eeq
\beq
\gamma_2=-\frac{g_2^2}{[(1+u)/2]^{D/2}}\Gamma(3-D/2).\label{gamma2}\eeq
\beq
\tau_1^a=-\frac{1}{4}g_1^2\Gamma(3-D/2),\label{taua}\eeq
\beq\tau_1^b=\frac{1}{4}\frac{g_3^2}{ u^{D/2}}(2-u)\Gamma(3-D/2),\label{taub}\eeq
\beq\tau_2=-\frac{g_2^2}{[(1+u)/2]^{D/2}}\frac{u}{1+u}\Gamma(3-D/2).\label{tau2}\eeq

Finally, we calculate $\kappa$.
\[\kappa_i=E_N\frac{\pd}{\pd E_N}\ln \Big(T_i^{-1}Z_i\Big)
=E_N\frac{\pd}{\pd E_N}\Big((Z_i-1)-(T_i-1)\Big).\]

From our expressions for $T_{i}$ we find
\[ E_N\frac{\pd}{\pd E_N}(T_{1}^a-1)=-g_1^2\Gamma(3-D/2),
\ \  E_N\frac{\pd}{\pd E_N}(T_{1}^b-1)=0,\]
\[ E_N\frac{\pd}{\pd E_N}(T_{2}-1)=-\frac{g_2^2}{[(1+u)/2]^{D/2}}\Gamma(3-D/2).\]
This gives
\beq
\kappa_1^a=\frac{1}{2}g_1^2\Gamma(3-D/2),
\eeq
\beq
\kappa_1^b=\frac{1}{2}\frac{g_3^2}{u^{D/2}}\Gamma(3-D/2),\eeq
\beq
\kappa_2=0.
\label{kappa2}
\eeq

To calculate $\beta$-functions one has to calculate the relevant diagrams for the non-trivial couplings.
In the single loop approximation which is our scope we have to calculate the adequate triangle diagrams
putting $\delta_1=\delta_2=0$. So the found $\beta$-functions are the same as in the massless case \cite{bkv2023}.
Here we reproduce them  (in the lowest order in small $\ep$).

At $u\neq 0$ the four $\beta$-functions are
\beq
\beta_1=-\frac{1}{4}\epsilon g_1+\frac{3}{2}g_1^3- g_2g_3^2\frac{2}{u^2}+g_1g_3^2\frac{1+u}{4u^2},
\label{bet1}
\eeq
\beq
\beta_2
=-\frac{1}{4}\epsilon g_2+g_1g_2^2\frac{6+2u}{(1+u)^2}- g_2g_3^2\frac{1+8u-u^2}{4u^2(1+u)},
\label{bet2}
\eeq
\beq
\beta_3
=-\frac{1}{4}\epsilon g_3+g_1g_2g_3\frac{4}{1+u}+g_2^2g_3\frac{4}{u(1+u)^2}+ g_3^3\frac{u-1}{4u^2},
\label{bet3}
\eeq
and
\beq
\beta_4=g_1^2\frac{u}{4}-g_2^2\frac{4u^2}{(1+u)^3}+ g_3^2\frac{u-2}{4u}.
\label{bet4}
\eeq

At $u=g_4=0$ one has to pass  from $g_3$ to a new coupling constant $r=g_3/g_4$
and the 4-dimensional domain of coupling constants $g_i$, $i=1,...,4$ splits into two 3-dimensional
domains with either $r=0$ or $g_2=0$.

If the initial $g_2=0$ then $\bar{g}_2(t)=0$ and $g_1,\ r$ and $g_4$ evolve with $\beta$-functions
\beq
\beta_1=-\frac{1}{4}\epsilon g_1+\frac{3}{2}g_1^3+g_1r^2\frac{1+u}{4},
\label{bet1b}
\eeq
\beq
\beta_r=r\Big(-\frac{1}{4}\ep- \frac{1}{4}g_1^2+ \frac{1}{4}r^2\Big)
\label{bettb}
\eeq
and
\beq
\beta_4=g_1^2\frac{u}{4}+ r^2\frac{u(u-2)}{4}.
\label{bet4b}
\eeq

If the intial $r=0$ then $\bar{r}(t)=0$ and $g_1,\ g_2$ and $g_4$ evolve with $\beta$-functions
\beq
\beta_1=-\frac{1}{4}\epsilon g_1+\frac{3}{2}g_1^3,
\label{bet1c}
\eeq
\beq
\beta_2
=-\frac{1}{4}\epsilon g_2+g_1g_2^2\frac{6+2u}{(1+u)^2},
\label{bet2c}
\eeq
\beq
\beta_4=g_1^2\frac{u}{4}-g_2^2\frac{4u^2}{(1+u)^3}.
\label{bet4c}
\eeq

The $\beta$-functions do not depend on masses $\delta_1$ and $\delta_2$. So the fixed points
can be borrowed from the study of the massless case. There are 5 real fixed points,
which are reproduced from  ~\cite{bkv2023} in Appendix 1.

\section{At the fixed point with small $\delta_1$}
\subsection{Scaling}
At the fixed point $g_i=g_{ic}$ we have
\beq\frac{dg_i(t)}{dt}=0,\ \  {\rm so\ that}\ \  \bg_i(t)=g_{ic}
\label{barg}\eeq
and is fixed during evolution together with $\gamma_i$. However, $\bar{E}$, $\balpha$ and $\bdel_2$ keep running
\beq
\bar{E}(-t)=Ee^{-t\zeta},\ \ \zeta=\frac{1}{1-\kappa_1(g_c)},\ \ \zeta-1=\frac{\kappa_1(g_c)}{1-\kappa_1(g_c)}
\eeq
\beq
\balpha'_1(-t)=\alpha'_1e^{-tz},\ \ z=\frac{1-\tau_1(g_c)}{1-\kappa_1(g_c)},\ \ z-1=-\frac{\tau_1(g_c)-\kappa_1(g_c)}{1-\kappa_1(g_c)}
\label{balpha}\eeq
and (with $\kappa_2=0$)
\beq
\bdel_2(-t)=\delta_2e^{-t\zeta}.\eeq
So
the solution (\ref{eq54}) at $g=g_c$ becomes
\[
\Gamma^R(E, k,g_c,\alpha'_1,\xi\delta_1,\delta_2,E_N)\]\beq=
\Gamma^R(Ee^{-t\zeta},k,g_c,\alpha'_1e^{-tz},\delta_1,\delta_2e^{-t\zeta},E_N)
e^{t[1-\sum_{i=1}^2(n_i+m_i)\gamma_i(g_c)/2]/[1-\kappa_1(g_c)]}.
\label{eq084}
\eeq

We use the scaling property
\[\Gamma^R(E,k,g_c,\alpha'_1,\delta_1,\delta_2,E_N)=
E_N\Big(\frac{E_N}{\alpha'_1}\Big)^{(2-n-m)D/4}\Phi\Big(\frac{E}{E_N},
\frac{\alpha'_1}{E_N}k^2,\frac{\delta_1}{E_N},\frac{\delta_2}{E_N},g_c\Big)\]
to obtain
\[\Gamma^R( E,k,g_c,\alpha'_1,\xi\delta_1,\delta_2,E_N)\]\[=
e^{t[1-\sum_{i=1}^2(n_i+m_i)\gamma_i(g_c)/2]/[1-\kappa_1(g_c)]}
E_N\Big(\frac{E_N}{\alpha_1'}\Big)^{(2-n-m)D/4} e^{tz(2-n-m)D/4}\]
\[
\times\Phi\Big(\frac{E}{E_N}e^{-t\zeta},\frac{\alpha'_1}{E_N}k^2e^{-tz},\frac{\delta_1}{E_N},\frac{\delta_2}{E_N}e^{-t\zeta},g_c\Big).\]
We denote
\[C(t)=e^{t[1-\sum_{i=1}^2(n_i+m_i)\gamma_i(g_c)/2]/[1-\kappa_1(g_c)]}e^{tz(2-n-m)D/4}.\]
Rescaling here $\delta_1\to \delta_1/\xi$ we get
\[\Gamma^R(E,k,g_c,\alpha'_1,\delta_1,\delta_2,E_N)=\]\[C(t)E_N\Big(\frac{E_N}{\alpha'_1}\Big)^{(2-n-m)D/4}
\Phi\Big(\frac{E}{E_N}e^{-t\zeta},\frac{\alpha'_1}{E_N}k^2e^{-tz},
\frac{\delta_1}{E_N\xi},\frac{\delta_2}{E_N}e^{-t\zeta},g_c\Big).\]
Taking
\[\xi=\frac{\delta_1}{E_N},\ \ t=\ln\frac{\delta_1}{E_N}\]
we find finally
\[\Gamma^R(E.k,g_c,\alpha'_1,\delta_1,\delta_2,E_N)=\]\[C(t)E_N\Big(\frac{E_N}{\alpha'_1}\Big)^{(2-n-m)D/4}
\Phi\Big(\frac{E}{E_N}e^{-t\zeta},\frac{\alpha'_1}{E_N}k^2e^{-tz},\frac{\delta_2}{E_N}e^{-t\zeta},g_c\Big).\]

In particular we find for the inverse full propagators
\beq
\Gamma_j(E,k^2,g_c,\alpha'_1,\delta_1,\delta_2,E_N)=E_N\Big(\frac{\delta_1}{E_N}\Big)^{[1-\gamma_j(g_c)]/[1-\kappa_1(g_c)]}
\Phi_j(\rho_1,\rho_2,\rho_3,g_c),
\ \ j=1,2.
\label{eq087}
\eeq
Here $\rho_i$ are given by
\beq
\rho_1=\frac{E}{E_N}e^{-t\zeta},\ \ \rho_2=\frac{\alpha'_1}{E_N}k^2e^{-tz},\ \ \rho_3=\frac{\delta_2}{E_N}e^{-t\zeta},
\label{rho}
\eeq
which can also be rewritten as
\beq
\rho_1=\frac{E}{\delta_1}\Big(\frac{\delta_1}{E_N}\Big)^{1-\zeta}\ \
,\ \ \rho_2=\frac{\alpha'_1k^2}{\delta_1}\Big(\frac{\delta_1}{E_N}\Big)^{1-z},\ \
\rho_3=\frac{\delta_2}{\delta_1}\Big(\frac{\delta_1}{E_N}\Big)^{1-\zeta}.
\label{rho1}
\eeq


\subsection{Scaling functions at the fixed point}
At the fixed point as $\epsilon\to 0$ constants $g_{1,2,3}^2$, $\gamma_i$ and $(Z-1)$ are proportional to $\epsilon$.
So the renormalized $\Gamma^R_i$ at $g=g_c$ are known in two first orders in the expansion in powers of $\epsilon$.
Comparing with its representation Eq.~(\ref{eq087}) in terms of the scaling function $\Phi_j(\rho_1.\rho_2,\rho_3)$, $j=1,2$
we can find the scaling functions $\Phi_j$  in the two first orders in $\epsilon$.
Suppressing for the moment subindex $j=1,2$ in $\Phi_j$ we have in these orders
\beq
\Phi(\rho_1,\rho_2,\rho_3)=\Phi_{0}(\rho_1,\rho_2,\rho_3)+\epsilon\Phi_{1}(\rho_1,\rho_2,\rho_3)+...\ ,
\label{phie}\eeq
where
$\Phi_0(\rho_1,\rho_2,\rho_3)=\Phi_{\ep=0}(\rho_1,\rho_2,\rho_3)$.
Note that only the form of $\Phi$ is taken at $\ep=0$ but not the arguments, which are also $\ep$-dependent.

We present
\[\frac{1-\gamma_i}{1-\kappa_1}=1-\tilde{\gamma_i},\ \
\tilde{\gamma}_i=\frac{\gamma_i-\kappa_1}{1-\kappa_1}.\]
At small $\ep$
\[\tilde{\gamma_i}=\gamma_i-\kappa_1\]

Calculations give (see Appendix 1) the following.

In the lowerst order in (\ref{phie}) for the pomeron and odderon we find
\beq
\Phi_{10}(\rho_1,\rho_2,\rho_3)=\rho_{1}-\rho_{2}-1,
\label{phi10}
\eeq
\beq
\Phi_{20}(\rho_1,\rho_2,\rho_3)=\rho_{1}-u\rho_{2}-\rho_{3}.
\label{phi20}
\eeq
In the linear order in $\ep$
\beq
\Phi_{11}(\rho_1,\rho_2,\rho_3)=-d_1x_1(\ln x_1-1)-d_3x_3(\ln x_3-1)+2d_3\rc(\ln(2\rc)-1),
\label{phi11}
\eeq
\beq
\Phi_{21}(\rho_1,\rho_2,\rho_3)=-d_2\Big[x_2(\ln x_2-1)+1\Big],
\label{phi21}
\eeq
where
\beq
x_1=\frac{1}{2}\rho_{2}+2-\rho_{1},\ \
x_2=\frac{u}{1+u}\rho_{2}+1+\rho_{3}-\rho_{1},\\ \
x_3=\frac{1}{2}u\rho_{2}+2\rho_{3}-\rho_{1}
\label{defxi1}
\eeq
and
the constants $d_i$ are defined from
\beq
\epsilon d_1=\frac{1}{2}g_1^2,\ \ \epsilon d_2=\frac{g_2^2}{[(1+u)/2]^{D/2}},\ \
\epsilon d_3=-\frac{g_3^2}{2u^{D/2}}.\label{defd123}\eeq

The logarithms in the expressions for the scaling functions
acquire imaginary parts $-i\pi$ when their arguments become negative, which happens at sufficiently
large values of energy. At $k^2=0$ this happens when $E>\min (2\delta_1,2\delta_2)$.

Note that in our derivation in Appendix 2  we actually constructed $\Phi$ with arguments $\rho_i$ taken at $\ep=0$.
 In fact $\Gamma_j$ will be expressed
by (\ref{eq087}) with the same function $\Phi(\rho_1,\rho_2,\rho_3)$ but of different arguments
 defined by (\ref{rho1}) and $\ep$-depending. So in the end we find  the inverse propagators Eq.~(\ref{eq087})
with $\rho$ from Eq.~(\ref{rho1}).

Actually, investigating the behaviour at $\delta_1\to 0$ one can safely put $\delta_2=0$. Indeed, if $\delta_2$
initially is greater than zero one can use perturbation treatment for diagrams with the odderon, so that the
only interesting case is when $\delta_2$ is exactly equal to zero. In our formulas this means that we may put
$\rc=\rho_3=0$. Then our scaling functions  $\Phi_{i}$ become simplified and we get
\beq
\Gamma_1(E,k^2,\alpha'_1,\delta_1,E_N)=\delta_1\Big(\frac{\delta_1}{E_N}\Big)^{-\tilde {\gamma}_1}
\Big\{\rho_1-\rho_2-1-\ep\Big(d_1x_1(\ln x_1-1)+d_3x_3(\ln x_3-1)\Big)\Big\}
\eeq
and
\beq
\Gamma_2(E,k^2,\alpha'_1,\delta_1,E_N)=\delta_1\Big(\frac{\delta_1}{E_N}\Big)^{-\tilde {\gamma}_2}
\Big\{\rho_1-u\rho_2-\ep\Big(d_2x_2(\ln x_2-1)+1\Big)\Big\},
\label{phi12a}
\eeq
where now ($\rho_i$ are defined in (\ref{rho}))
\beq
x_1=\frac{1}{2}\rho_{2}+2-\rho_{1},\ \ x_2=\frac{u}{1+u}\rho_{2}+1-\rho_{1},\ \
x_3=\frac{1}{2}u\rho_{2}-\rho_{1}.
\label{defxia}
\eeq

The actual behaviour at $\delta_1\to 0$ depends on the parameters in these
formulas. These parameters depend on the choice of fixed points.
As indicated in Appendix 1 the only purely attractive fixed point is $g_c^{(3)}$, for which at
$D=2$ we find parameters
\[\tilde{\gamma}_1=-\frac{2}{5},\ \ \tilde{\gamma}_2=-\frac{3}{10},\ \ \zeta=\frac{6}{5},\ \ z=\frac{13}{10},
\ \ d_1=2d_2=\frac{1}{12},\ \ d_3=0.\]
Note that at the fixed point $g_c^{(1)}$ parameters $\tilde{\gamma}_1$, $\zeta$ and $z$ are the same, the only difference
being in $\tilde{\gamma}_2$.  Without odderons $g_c^{(1)}$ becomes attractive. It means that inclusion of odderons does not change the
behaviour of $\Gamma_1$ at least at purely attractive fixed points (different with or without odderons)
\footnote{ The asymptotical $\Gamma_1$ at small $\delta_1$ was calculated without odderons in  ~\cite{abarb2}.
To compare one has to note that the parameters were taken there in the limit $\ep\to 0$ up to linear terms.}.

The actual asymptotic at $\delta_1\to 0$ is determined by the fact that according to (\ref{rho1}) with $\zeta$ and $z$ greater than
unity $\rho_1$ and $\rho_2$ infinitely grow as $\delta_1\to 0$ and with them also $x_i$.
As a result, the limiting expressions come from the logarithmic term in
$\Phi_1$. Taking for simplicity $k^2=0$ and so $\rho_2=0$ one gets in this limit
\beq
\Gamma_1=\frac{1}{5}E\Big(\frac{\delta_1}{E_N}\Big)^{1/5}\ln\frac{E}{\delta_1},
\ \ \Gamma_2=\frac{1}{10}E\Big(\frac{\delta_1}{E_N}\Big)^{1/10}\ln\frac{E}{\delta_1}.
\eeq

\subsection{Trajectories}
The inverse propagators $\Gamma_i$, $i=1,2$  have each a zero at some point  at which
\[\Phi_i(\rho_1,\rho_2,\rho_3,g_c)=0.\]
Consider $\Phi_i$ at fixed $\rho_3$
\[\Psi(\rho_1,\rho_2)\equiv\Phi_i(\rho_1,\rho_2,\rho_3).\]
Then we can proceed as in ~\cite{abarb2}.
Let the zero of $\Psi$ occur at $\rho_{1c}$ when $\rho_2=0$.
Of course, now $\rho_{1c}$ depends not only on $g_c$ but also on $\delta_2$.
Expanding $\Psi(\rho_1,\rho_2)$ in small $\rho_2$ around this point we find
\[\frac{\pd\Psi}{\pd\rho_1}(\rho_1-\rho_{1c})+\frac{\pd \Psi}{\pd\rho_2}\rho_2=0,\]
where the derivatives are taken at $\rho_1=\rho_{1c}$ and $\rho_2=0$.
From this, remembering that $E=1-\alpha(t=-k^2)$,
one finds the trajectories for the pomeron and odderon
(indices $i=1,2$ are suppressed)
\beq
\Delta=1-\alpha(0)=\delta_1\Big(\frac{\delta_1}{E_N}\Big)^{\zeta-1} \rho_{1c},\label{ddelta}\eeq
\beq
\alpha_R'(0)=\Big(\frac{\delta_1}{E_N}\Big)^{\zeta-z}\alpha'_{1}R,\ \ R=-\frac{\pd \Psi}{\pd\rho_2}\Big/\frac{\pd \Psi}{\pd\rho_1}.
\label{alpha}
\eeq

Knowing $\Phi$ one can determine the trajectories in the explicit form using (\ref{ddelta}) and (\ref{alpha}).

For the pomeron  the equation to calculate the point $\rho_{1c}$ is
$\Phi_1(\rho_{1c},0,\rho_3)=0$.
In the lowest approximation from (\ref{phi10}) we get $\rho_{1c}^{(0)}=1$. In the next order
\[\rho_{1c}-1+\ep\Phi_{11}(\rho_{1c},0,\rho_3)=0.\]
Solving it up to terms linear in $\ep$ we find
\[\rho_{1c}=1-\ep\Phi_{11}(\rho_{1c}^{(0)},0,\rho_3),\]
or explicitly
\beq
\rho_{1c}=1-\ep\Big[d_1-d_3(2\rho_3-1)\Big(\ln(2\rho_3-1)-1\Big)+2d_3\rho_3\Big(\ln (2\rho_3)-1\Big)\Big],
\label{r1c}
\eeq
where
\[\rho_3=\frac{\delta_2}{\delta_1}\Big(\frac{\delta_1}{E_N}\Big)^{1-\zeta}.\]

According to (\ref{alpha}) to find the slope one has to calculate derivatives of $\Phi$ in $\rho_1$ and $\rho_2$.
We have
\[\frac{\pd \Phi_{10}}{\pd \rho_1}=1,\ \ \frac{\pd \Phi_{10}}{\pd \rho_2}=-1,\]
\[\frac{\pd \Phi_{11}}{\pd \rho_1}=-\frac{\pd \Phi_{11}}{\pd x_1}-\frac{\pd \Phi_{11}}{\pd x_2}-\frac{\pd \Phi_{11}}{\pd x_3},\]
\[\frac{\pd \Phi_{11}}{\pd \rho_2}=\frac{1}{2}\frac{\pd \Phi_{11}}{\pd x_1}+\frac{u}{1+u}\frac{\pd \Phi_{11}}{\pd x_2}
+\frac{u}{2}\frac{\pd \Phi_{11}}{\pd x_3}.\]
Finally,
\[\frac{\pd \Phi_{11}}{\pd x_1}=-d_1\ln x_1,\ \ \frac{\pd \Phi_{11}}{\pd x_2}=0,\ \
\frac{\pd \Phi_{11}}{\pd x_3}=-d_3\ln x_3.\]
So we find
\[
\frac{\pd \Phi_1}{\pd \rho_1}=1+\ep d_3\ln x_3,\ \
\frac{\pd \Phi_{1}}{\pd \rho_2}=-1-\ep\frac{u}{2}d_3\ln x_3,\]
where
$x_3=2\rho_3-1$.
The ratio $R$ in (\ref{alpha}) turns out to be
\beq
R_1=- \frac{\pd\Phi_{1}}{\pd \rho_2}\Big/ \frac{\pd\Phi_{1}}{\pd \rho_1}
=\frac{1+\ep ud_3\ln x_3/2}{1+\ep d_3\ln x_3}\simeq 1+\ep\Big(\frac{u}{2}-1\Big)d_3\ln x_3.
\label{rat1}
\eeq
The final values for the intercepts and slope is given by Eqs.~(\ref{ddelta}) and (\ref{alpha}) with
$\rho_{1c}$ and the ratio of derivatives given by
(\ref{r1c}) and (\ref{rat1}).

Note that at the purely attractive fixed point $g_c^{(3)}$ coefficient $d_3=0$. So  at this point
\[\Delta=\frac{5}{6}\delta_1\Big(\frac{\delta_1}{E_N}\Big)^{1/5},\]
Also we find in this case $R_1=1$, so both intercept and slope do not depend on $\delta_2$
and are the same as obtained without odderons in ~\cite{abarb2}.

Passing to the odderon we similarly have
the equation
$\Phi_2(\rho_{2c},0,\rho_3)=0$.
In the lowest approximation from (\ref{phi20}) we get $\rho_{2c}^{(0)}=\rho_3$. So in the next order
\[\rho_{2c}-\rho_3+\ep\Phi_{21}(\rho_3,0,\rho_3)=0.\]
We have
\[\Phi_{21}(\rho_3,0,\rho_3)=-d_2\Big[x_2(\ln x_2-1)+1\Big],\]
where
\[x_2=\Big(\frac{u}{1+u}\rho_2+1+\rho_3-\rho_1\Big)_{\rho_1=\rho_3,\rho_2=0}=1,\]
so that $\Phi_{21}(\rho_3,0,\rho_3)=0$ and
$\rho_{2c}=\rho_3$.%

Now we go the slope.
We have in the lowest approximation
\[\frac{\pd\Phi_{20}}{\pd\rho_1}=1,\ \ \frac{\pd\Phi_{20}}{\pd\rho_2}=-u.\]
In the second order we find
\[\frac{\pd\Phi_{21}}{\pd x_1}=\frac{\pd\Phi_{21}}{\pd x_3}=0,\ \
\frac{\pd\Phi_{21}}{\pd x_2}=-d_2\ln x_2.\]
Since $x_2=1$, also
\[\frac{\pd\Phi_{21}}{\pd x_2}=0.\]
As a result,
\[\frac{\pd\Phi_{2}}{\pd \rho_1}=1,\ \ \frac{\pd\Phi_{2}}{\pd \rho_2}=-u.\]
We find the ratio $R$ in (\ref{alpha})
\beq
R_2=- \frac{\pd\Phi_{2}}{\pd \rho_2}\Big/ \frac{\pd\Phi_{2}}{\pd \rho_1}=u.
\label{rat2}
\eeq
One has
\[E_N\rho_3\Big(\frac{\delta_1}{E_N}\Big)^\zeta=\delta_2.\]
So the intercept of the odderon trajectory does not change
with the interaction, whereas its slope changes and depends
on $\delta_1$:
\beq
\Delta_2=\delta_2,
\ \ {\alpha'}^R_2=\Big(\frac{\delta_1}{E_N}\Big)^{\zeta-z}u\alpha'_1
=\Big(\frac{\delta_1}{E_N}\Big)^{\zeta-z}\alpha'_2.
\label{trajod}
\eeq
At the purely attractive fixed point $g_c^{(3)}$ constant $g_4\equiv u=0$
(see Table~1) and for the finite value of parameter $\alpha'_1$
one meets the limit case of ''flat trajectory'' with the intercept
$\Delta_2=\delta_2$ and a zero slope ${\alpha'}^R_2=\alpha'_2=0$.



\section{Small $\delta_2$}
\subsection{Scaling functions}
In the previous sections 2 and 4 we studied the behaviour of the generalized vertices when $\delta_1$ is small:
$\delta_1\to \xi\delta_1$ and $\delta_1\to 0$.
In this section in the similar way we study the behaviour at $\delta_2\to 0$.

Substituting in Eq.~(\ref{gamxi}) $\delta_2$ by $\xi\delta_2$  instead of (\ref{eq51}) we get
\beq
\Gamma^{R}(E,k^2,g,\alpha',\delta_1,\xi\delta_2,E_N)=
\xi\Gamma^{R}\Big(\frac{E}{\xi},k^2,g,\frac{\alpha'}{\xi},\frac{\delta_1}{\xi},\delta_2,\frac{E_N}{\xi}\Big).
\label{eq51a}
\eeq
Next derivation follows  the one which lead from Eq.~(\ref{eq51}) to  the solution (\ref{eq54}).
Note, however, that in the single loop approximation $\kappa_2=0$, which simplifies evolution equations. Putting
$t=\ln\xi$ we obtain
\[
\Gamma^R(E,k^2,g,\alpha'_1,\delta_1,\xi\delta_2,E_N)=
\Gamma^R\Big(\bar{E}(-t),k^2,\bar{g}(-t),\bar{\alpha}'_1(-t),\bar{\delta}_1(-t),\delta_2,E_N\Big)\]
\beq
\times
\exp\Big\{\int_{-t}^0dt'[1-\gamma(g_1(t'))]\Big\},
\label{eq54a}
\eeq
where
\[
\frac{d \bar{g_i}(t)}{dt}=-\beta_i(\bar{g}(t)),\]
\[\frac{d\ln\bar{\alpha}'_1(t)}{dt}=1-\tau_1(\bar{g}(t)),\]
\[\frac{d\ln\bar{\delta}_1(t)}{dt}=1-\kappa_1(\bar{g}(t)),\]
\beq
\frac{d\ln\bar{E}(t)}{dt}=1
\label{eq55a}
\eeq
with the initial conditions
\[\bar{g_i}(0)=g_i,\ \ \bar{\alpha}_1'(0)=\alpha'_1,\ \ \bar{\delta}_1(0)=\delta_1,\ \ \bar{E}(0)=E.\]

Next  changes concern the content of section 4 dealing with the  situation at  fixed points.
Now  the running parameters are
\beq
\bar{E}(-t)=Ee^{-t},
\eeq
\beq
\balpha'_1(-t)=\alpha'_1e^{-tz},\ \ z=1-\tau_1(g_c)
\label{balphaa}\eeq
and
\beq
\bdel_1(-t)=\delta_1e^{-t\zeta},\ \  \zeta=1-\kappa_1(g_c)\eeq
Solution (\ref{eq54a}) at $g=g_c$ becomes
\[
\Gamma^R(E, k,g_c,\alpha'_1,\delta_1,\xi\delta_2,E_N)\]\beq=
\Gamma^R(Ee^{-t},k,g_c,\alpha'_1e^{-tz},\delta_1e^{-t\zeta},\delta_2,E_N)
e^{t[1-\sum_{i=1}^2(n_i+m_i)\gamma_i(g_c)/2]}.
\label{eq084a}
\eeq
As before we use the scaling property with
$\delta_2\to \delta_2/\xi$.
Taking
\[\xi=\frac{\delta_2}{E_N},\ \ t=\ln\xi\]
we find
\[\Gamma^R(E,k,g_c,\alpha'_1,\delta_1,\delta_2,E_N)=\]\[C(t)E_N\Big(\frac{E_N}{\alpha'_1}\Big)^{(2-n-m)D/4}
\Phi\Big(\frac{E}{E_N}e^{-t},\frac{\alpha'_1}{E_N}k^2e^{-tz},
\frac{\delta_1}{E_N}e^{-t\zeta},g_c\Big),\]
where
\[C(t)=e^{t[1-\sum_{i=1}^2(n_i+m_i)\gamma_i(g_c)/2]}e^{tz(2-n-m)D/4}.\]

In particular, we find for the inverse full propagators
\beq
\Gamma_j(E,k^2,g_c,\alpha'_1,\delta_1,\delta_2,E_N)=
\delta_2\Big(\frac{\delta_2}{E_N}\Big)^{-\gamma_j(g_c)}
\Phi_j(\rho_1,\rho_2,\rho_3,g_c),
\ \ j=1,2,
\label{eq087a}
\eeq
where
\beq
\rho_1=\frac{E}{\delta_2},\ \ \rho_2=\frac{\alpha'_1}{E_N}k^2e^{-tz},\ \ \rho_3=\frac{\delta_1}{E_N}e^{-t\zeta}.
\label{rho2}
\eeq

Calculation of the scaling functions repeats the previous one for $\delta_1\to 0$. It can  be found in Appendix 2.
As a result, we get
\beq
\Phi_{10}(\rho_{1},\rho_{2},\rho_{3})=\rho_{1}-\rho_{2}-\rho_{3},
\label{phi10a}
\eeq
\beq
\Phi_{20}(\rho_{1},\rho_{2},\rho_{3})=\rho_{1}-u\rho_{2}-1
\label{phi20a}
\eeq
and in the linear order in $\ep$
\beq
\Phi_{11}(\rho_1,\rho_2,\rho_3)=
-d_1x_1(\ln x_1-1)-d_3x_3(\ln x_3-1)
+2d_3(\ln(2)-1)
\label{phi11a}
\eeq
and
\beq\Phi_{21}(\rho_1,\rho_2,\rho_3)
=d_2\Big[-x_2(\ln x_2-1)
+\rc(\ln\rc-1)\Big]
\label{phi21a}
\eeq
with $x_i$ slightly different from (\ref{defxi1})
\beq
x_1=\frac{1}{2}\rho_{2}+2\rho_{3}-\rho_{1},\ \ x_2=\frac{u}{1+u}\rho_{2}+1+\rho_{3}-\rho_{1},\ \
x_3=\frac{1}{2}u\rho_{2}+2-\rho_{1} .
\label{defxi2}
\eeq
As in the case when $\delta_1\to 0$, in this case, investigating the behaviour at $\delta_2\to 0$,
we can put $\delta_1=0$
and so set $\rho_3=0$. This simplifies our $\Phi_{21}$ giving
\beq\Phi_{21}(\rho_1,\rho_2,0)
=-d_2x_2(\ln x_2-1).
\eeq
Then we get
\[
\Gamma_1(E,k^2,\alpha'_1,\delta_1,E_N)=\delta_2\Big(\frac{\delta_2}{E_N}\Big)^{-\gamma_1}\]\beq\times
\Big\{\rho_1-\rho_2+\ep\Big(-d_1x_1(\ln x_1-1)-d_3x_3(\ln x_3-1)
+2d_3(\ln(2)-1)\Big)\Big\},
\label{gam1b}\eeq
\beq
\Gamma_2(E,k^2,\alpha'_1,\delta_1,E_N)=\delta_2\Big(\frac{\delta_2}{E_N}\Big)^{-\gamma_2}
\Big\{\rho_{1}-u\rho_{2}-1-\ep d_2x_2(\ln x_2-1)\Big\}.
\label{gam2b}\eeq
In these expressions one has to take $\rho_3=0$ in $x_i$
\beq
x_1=\frac{1}{2}\rho_{2}-\rho_{1},\ \ x_2=\frac{u}{1+u}\rho_{2}+1-\rho_{1},\ \
x_3=\frac{1}{2}u\rho_{2}+2-\rho_{1} .
\label{defxi2b}
\eeq

The actual behaviour at $\delta_2\to 0$ as before depends on the parameters, which in their turn
 depend on the choice of fixed points.
For the only purely attractive fixed point is $g_c^{(3)}$  at
$D=2$ we find parameters
\[\gamma_1=-\frac{1}{6},\ \ \gamma_2=-\frac{1}{12},\ \ \zeta=\frac{5}{6},\ \ z=\frac{13}{12}.\]

The asymptotical behaviour at $\delta_2\to 0$ is similar to that for $\delta_1\to 0$ and, since $\rho_i$ grow,
comes from the logarithmic terms in $\Phi_1$. It becomes especially simple if we put $k^2=0$ as before. Then
$\rho_1=E/\delta_2$ and the asymptotic is determined by the $\gamma$  in the exponent
\[\Gamma_1=\frac{1}{6}E\Big(\frac{\delta_2}{E_N}\Big)^{1/6}\ln\frac{E}{\delta_2},
\ \ \Gamma_2=\frac{1}{12}E\Big(\frac{\delta_2}{E_N}\Big)^{1/12}\ln\frac{E}{\delta_2}.\]


\subsection{Trajectories}
The trajectories are calculated using the same expressions as for $\delta_1\to 0$
except that we are to put $\zeta=1$ in Eqs.~(\ref{ddelta}) and (\ref{alpha}) due to the simple evolution of $E$
\beq
\Delta=1-\alpha(0)=\delta_2 \rho_{1c},
\label{ddelta2}\eeq
\beq
\alpha_R'(0)=\Big(\frac{\delta_2}{E_N}\Big)^{1-z}\alpha'_{1}R,\ \ R=-\frac{\pd \Psi}{\pd\rho_2}\Big/\frac{\pd \Psi}{\pd\rho_1}.
\label{alpha2}
\eeq

For the pomeron in the lowest approximation we
obviously find $\rho_{1c}=\rho_3$, so that in the next
approximation we have
\[\rho_{1c}=\rc-\ep\Phi_{11}(\rho_3,0,\rho_3).\]
The scaling function $\Phi_{11}$ is given in Eq.~(\ref{phi11a})
with $x_1$ and $x_3$, which take values
\[x_1=\rho_3,\ \ x_3=2-\rho_3,\]
where $\rho_3$ is given by (\ref{rho2}).

Passing to the slope we find the necessary derivatives
\[
\frac{\pd\Phi_1}{\pd \rho_1}=1+\ep\Big(d_1\ln\rho_3+d_3\ln(2-\rho_3)\Big),\]
\[
\frac{\pd\Phi_1}{\pd \rho_2}=-1-\ep\frac{1}{2}
\Big(d_1\ln\rho_3+d_3u\ln(2-\rho_3)\Big).\]
The ratio of derivatives in (\ref{alpha2}) is up to linear terms in $\ep$
\beq
R_1=1-\ep\frac{1}{2}\Big(d_1\ln\rho_3+(2-u)d_3\ln(2-\rho_3)\Big).
\label{r1a}
\eeq

For the attractive fixed point $g_c^{(3)}$ we have $d_3=0$ and so
\[\Delta=\delta_2\rho_3\Big(1+\frac{1}{6}(\ln \rho_3-1)\Big),\]
\[R_1=1-\frac{1}{12}\ln\rho_3.\]
Both depend on $\rho_3$ and thus on $\delta_1$.  So at $\delta_2\to 0$ at this
fixed point the pomeron trajectory  depends on the fixed $\delta_1$, contrary to what occurs at $\delta_1\to 0$.

For the odderon in the lowest approximation we get
$\rho_{2c}=1$. In the next approximation we obtain
\[\rho_{2c}=1-\ep\Phi_{21}(1,0,\rc).\]
In $\Phi_{21}$ given by (\ref{phi21a}) we have to put
$x_2=\rho_{30}$
and the  two terms cancel.  We find
\[\Delta=\delta_2,\]
so that the odderon intercept does not depend on $\delta_1$ and remains trivial.

The necessary derivatives are found to be
\[
\frac{\pd\Phi_2}{\pd \rho_1}=1+\ep d_2\ln\rc,\ \
\frac{\pd\Phi_2}{\pd \rho_2}=-u-\ep \frac{u}{1+u}d_2\ln\rc,\]
so that the ratio of interest is (up to terms linear in $\ep$)
\[
R_2=u\Big(1-\ep d_2\frac{u}{1+u}\ln\rc\Big),
\]
which means that the slope changes 
(as in the case $\delta_1=0$)
\beq
{\alpha'}^R_2=\Big(\frac{\delta_2}{E_N}\Big)^{1-z}
\Big(1-\ep d_2\frac{u}{1+u}\ln\frac{\delta_1}{\delta_2}\Big)\alpha'_{2} .
\label{r2a}
\eeq
At the attractive fixed point $g_c^{(3)}$ we also have the odderon
trajectory with a zero slope ${\alpha'}^R_2=\alpha'_2=0$.


\section{Elastic scattering amplitude}
\subsection{The asymptotic }
We consider the elastic scattering of two particles with exchanges of  pomerons and odderons.
It is the sum of contributions in which the projectile emits $n_1$ pomerons and $m_1$ odderons and the target absorbs
$n_2$ pomerons and $m_2$ odderons
\[{\cal A}(s,t)=\sum_{(n,m)}{\cal A}^{(nm)}(s,t).\]
Here $(nm)=(n_1,n_2,m_1,m_2)$ where $n_1$ and $n_2$ are numbers of incoming pomerons and odderons and
$m_1$ and $m_2$ are numbers of outgoing pomerons and odderons. In the following we denote
the number of initial reggeons (pomeron plus odderons) $n=n_1+n_2$, the number of final reggeons $m=m_1+m_2$, the total number of pomerons
(initial plus final) $n_P=n_1+m_1$, the total number of odderons $n_O=n_2+m_2$. The total number of all reggeons is evidently
$n_t=n+m=n_P+n_O$.
Amplitude ${\cal A}$ with a given number of exchanged reggeons is shown in Fig.~2.
\begin{figure}
\begin{center}
\epsfig{file=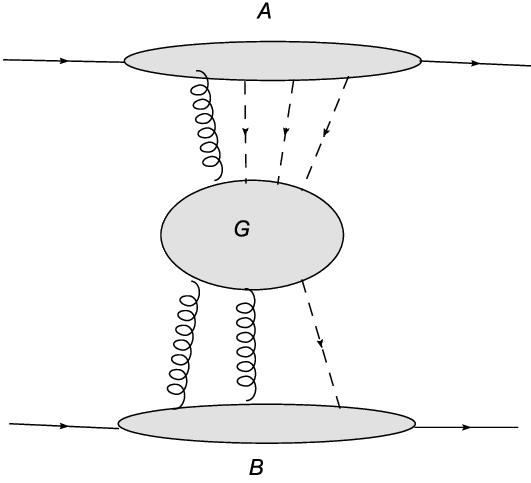, width=8 cm}
\caption{Elastic amplitude with a given number of pomerons (solid lines) and odderons (dashed lines) exchanges.}
\end{center}
\label{fig2}
\end{figure}
For simplicity we assume that the couplings of the reggeons  to the participants are just (unknown) constants,
namely $A^{n_1,m_1}$ and $B^{n_2,m_2}$.
This roughly speaking corresponds to the Glauber coupling. In this case the Mellin-transformed amplitude, that is
in the complex angular momentum space variables, will be given by the integral over all internal energetic and momenum
integration variables
\[
{\cal A}^{(nm)}(E,q)=A^{n_1,n_2}B^{m_1,m_2} I^{(nm)}(E,t),\]
where
\[
I^{(nm)}(E,t)=\]\beq=\int\prod_{i=1}^{n+m}d^Dk_idE_i
\delta^D(\sum_{in}k_i-q)\delta^D(\sum_{out}k_i-q)
\delta(\sum_{in}E_i-E)\delta(\sum_{out}E_i-E)
G_R^{(nm)}(E_i,k_i)
\label{eq0110}
\eeq
and $t=-q^2$ is the total transferred momentum squared.
Summations inside $\delta$-functions go over energies and momenta of  the incoming and outgoing reggeons.

The full Green function $G^{(nm)}$ is a product of the amputated one and $n+m$ propagators, that is the inverse
$\Gamma_R^{1,0,1,0}\equiv \Gamma_R^{(1)}$ for the pomerons and $\Gamma_R^{0,1,0,1}\equiv \Gamma_R^{(2)}$  for the odderons.

Let us start with the case when the Green function does not contain  disconnected parts. Then
\beq
G_R^{(nm)}(E_i,k_i)=\Gamma_R^{(nm)}(E_i,k_i)
\prod_{i=1}^{n_P}\Big(\Gamma_R^{(1)}(E_i,k_i)\Big)^{-1}\prod_{i=1}^{n_O}\Big(\Gamma_R^{(2)}(E_i,k_i)\Big)^{-1},
\label{eqad1}\eeq
where $\Gamma_R$ are connected amputated Green functions considered previously.

Our aim is to use the scaling properies of $G_R$ in the integrand. For simplicity we shall consider the simpler case when $\delta_1=\delta_2=0$
so that the model formally becomes identical with the one without masses studied in ~\cite{bkv2023}. This allows to use the scaling properties
established in that publication. Namely

\[\Gamma_R^{(nm)}(E_i,k_i,g_c,\alpha'_1,E_N)=E_N\Big(\frac{E_N}{\alpha'_1}\Big)^{(2-n-m)D/4}
\xi^{1-\sum_{i=1}^2(n_i+m_i)\gamma_i(g_c)/2+z(2-n-m)D/4}\]\beq\times
\Phi^{(nm)}\Big(\frac{-E_i}{E},\xi^{-z}\frac{{k}_i{k}_j}{E_N}\alpha'_1,g_c\Big),\label{gamrphi}\eeq
where $\gamma_1$, $\gamma_2$ and $z=1-\tau_1(g_c)$ are the anomalous dimensions.
In particular,
\beq
\Gamma_R^{(i)}(E,k^2,g_c,\alpha'_1,E_N)=E_N\xi^{1-\gamma_i(g_c)}\Phi_{i}\Big(\xi^{-z}\frac{\alpha'_1k^2}{E_N},g_c\Big),
\ \ i=1,2.
\label{eq087b}
\eeq
In these formulas
\[\xi=\frac{-E}{E_N},\]
where $E_N$ is the renormalization energy.
Putting (\ref{gamrphi}) and (\ref{eq087b}) in (\ref{eqad1}) we find the scaling properties of $G_R^{(nm)}$
\[
G_R^{(nm)}(E_i,k_i)=E_N^{1-n_t}\Big(\frac{E_N}{\alpha'_1}\Big)^{(2-n_t)D/4}\xi^c
\Phi^{(nm)}\Big(-\frac{E_i}{E},\xi^{-z}\frac{{k}_i{k}_j}{E_N}\alpha'_1,g_c\Big)\]
\beq
\prod_{i=1}^{n_P}\Big[\Phi_{1}\Big(\xi^{-z}\frac{k_i^2}{E_N},g_c\Big)\Big]^{-1}
\prod_{i=1}^{n_O}\Big[\Phi_{2}\Big(\xi^{-z}\frac{k_i^2}{E_N},g_c\Big)\Big]^{-1},
\label{eqad2}\eeq
where
\[c=1-n_t+\frac{1}{2}\gamma_1 n_P+\frac{1}{2}\gamma_2n_O+z(2-n_t)\frac{D}{4}.\]

To extract the total dependence of $I^{(nm)}(E,t)$ we make a change of integration variables
\[E_i=E\zeta_i,\ \ {k}_i=\xi^{z/2}{x}_i.\]
This change gives an extra factor
\[E^{n_t-2}\xi^{z(n_t-2)D/2}\]
and the $\delta$-functions turn into
\[\delta(\sum \zeta_i-1)\ \ {\rm and}\ \ \delta^D(\sum x_i-q\xi^{-z/2})\]
for integrations over incoming and outgoing energies and momenta.

In the end we get
\beq
I^{(nm)}(E,t)=E^{-1+a}F^{(nm)}(t\xi^{-z}),
\label{eqad3}
\eeq
where
\beq a=\frac{1}{2}\gamma_1n_P+\frac{1}{2}\gamma_2n_O+\frac{1}{4}zD(n_t-2)
\label{eqad4}\eeq
and some functions  $F(t\xi^{-z)})$, which
are determined  by functions $\Phi$ including also factors from the definition of $\xi$ in terms of $E$.

The amplitude is obtained as the inverse Mellin transform. For given $(nm)$
\[
{\cal A}^{(nm)}(s,t)=A^{n_1,m_1}B^{n_2,m_2}\frac{s}{2\pi i}\int_{\sigma-i\infty}^{\sigma+i\infty}dEe^{-Ey}  I^{(nm)}(E,t)\]\beq
=\frac{s}{2\pi i}\int_{\sigma-i\infty}^{\sigma+i\infty}\frac{dE}{E}e^{-Ey}E^a\tilde{F}(t(-E)^{-z}E_N^z),
\label{eqad5}\eeq
where $y=\ln s$ and $\tilde{F}$ includes the impact factors $A$ and $B$.
Changing integration variables $Ey=\varepsilon$ we get

\[{\cal A}^{(nm)}(s,t)=sy^{-a}\frac{1}{2\pi i}\int_{\sigma'-i\infty}^{\sigma'+i\infty}
\frac{d\varepsilon}{\varepsilon}\varepsilon^ae^{-\varepsilon}\tilde {F}\Big(t y^z \Big(\frac{-\epsilon}{E_N}\Big)^{-z}\Big).\]
Denoting the result of integration over $\varepsilon$ as $\Psi(ty^z)$ we find our final result
\beq
{\cal A}^{(nm)}(s,t)=sy^{p(n_p,n_o)}\Psi(ty^z),
\label{eqad6}\eeq
where the power $p=-a$ is
\beq
p(n_P,n_O)=z\frac{D}{2}-n_P\Big(\frac{1}{2}\gamma_1+z\frac{D}{4}\Big)-n_O\Big(\frac{1}{2}\gamma_2+z\frac{D}{4}\Big).
\label{eqad7}
\eeq

We take $D=2$. Then for the simplest exchanges we get:\\
one pomeron exchange
\[p(2,0)=-\gamma_1,\]
one odderon exchange
\[p(0,2)=-\gamma_2.\]
Exchange by one more pomeron gives the change of power
\[\Delta_P=p(n_P+1,n_O)-p(n_P,n_O)=-\frac{1}{2}(\gamma_1+z).\]
Exchange by two more odderons gives the change of power
\[\Delta_O=p(n_P,n_O+2)-p(n_P,n_O)=-(\gamma_2+z).\]

One can show that these results do not change if the Green function contains disconnected parts (see Appendix 3).

The further  study of the asymptotical behaviour (\ref{eqad6}) depends on the numerical values of the anomalous dimensions
at different fixed points.

%

\subsection{At fixed point and  with $D=2$}
Values of $\gamma_1$, $\gamma_2$, $\tau_1$ and $z$ for the five found real points are shown in Table 6 in Appendix 1.
From these values at $D=2$ we find for different fixed points

\[g_c^{(0)}:\ \ \ p(2,0)=-1,\ \ p(0,2)=0, \
\Delta_P=-\frac{1}{2},\ \ \Delta_O=0.\]

\[g_c^{(1)}:\ \ \ p(2,0)=\frac{1}{6},\ \ p(0,2)=0,\ \
\Delta_P=-\frac{11}{24},\ \ \Delta_O=-\frac{13}{12}.\]

\[g_c^{(2)}:\ \ \ p(2,0)=\frac{1}{6},\ \ p(0,2)=\frac{2}{11.3},\ \
\Delta_P=-\frac{11}{24},\ \ \Delta_O=-\Big(\frac{13}{12}-\frac{2}{11.3}\Big).\]

\[g_c^{(3)}:\ \ \ p(2,0)=\frac{1}{6},\ \ p(0,2)=\frac{1}{12},\ \
\Delta_P=-\frac{11}{24},\ \ \Delta_O=-1.\]

\[g_c^{(4)}:\ \ \ p(2,0)=-1,\ \ p(0,2)=0,\ \
\Delta_P=-1,\ \ \Delta_O=-1.\]

Inspecting these results we find the following.

$\bullet$ All $\Delta_P$ are negative. So the leading contribution comes from
the minimal number of exchanged pomerons.

$\bullet$ For all fixed points except $g_c^{(0)}$ also $\Delta_O$ is negative,
so that the leading contribution comes from the minimal number of exchanged odderons.
At the fixed points $g_c^{(0)}$ we find $\Delta_O=0$ and the asymptotic is the same for any number of exchanged odderons.

$\bullet$ At $g_c^{(1,2,3)}$ the cross-sections due to the single pomeron exchange grow as $y^{1/6}$.
At $g_c^{(0,4)}$ the cross-sections fall as $1/y$.

$\bullet$ At $g_c^{(0,1,4)}$ the cross-sections due to the single odderon exchange are constant.
At $g_c^{(2)}$ the cross-section rises roughly as $y^{1/6}$.
Notably at $g_c^{(3)}$ it rises as $y^{1/12}$, however, not so fast as the one-pomeron exchange ($\sim y^{1/6}$).

$\bullet$ So finally in all important cases when the single pomeron contribution grows it dominates over all multireggeon
contributions, as in absence of odderons, which result was found in ~\cite{abarb1}.

Taking into account that the only totally attractive fixed point is $g_c^{(3)}$ we conclude
from our study that
most probably the leading contribution will be the single pomeron exchange and the subdominant one the single odderon exchange
\beq
{\cal A}(s,t)=sy^{1/6}\Psi_{20}(ty^{13/12})+sy^{1/12}\Psi_{02}(ty^{13/12})
\label{eqad10}\eeq
with the cross-section of the form
\beq
\sigma^{tot}=y^{1/6}A+y^{1/12}B+{\cal O}(y^{-7/24}) .
\label{eqad11}\eeq

\section{Conclusions}
Using the renormalization group technique we studied the Regge model with the pomeron and odderon
interacting with triple vertices and imaginary coupling constants at different masses $\delta_1$ and $\delta_2$
for the pomeron and odderon respectively. The  masses in the renormalized Lagrangian are originally positive
and turn to zero as the physical intercept with all interactions included goes to unity.
Our primary goal has been to find the behaviour of observable in the limit $\delta_{1,2}\to 0$
with a view do see presence of a singularity at that point, which would indicate transition to
a new phase, which in all probability would be non-physical due to violation of the
projectile-target symmetry. So the presence of singularity means the model cannot be correctly defined
for negative $\delta_{1,2}$, or the supercritical pomeron and odderon.

As a rule at fixed points $\beta_i=0$, $i=1,...,4$ the singularity in question turns out to be a branch point
$\delta_j^{\tilde{\gamma}_j}$, $j=1,2$ with non-integer $\tilde{\gamma}_{1,2}$
at both the critical dimension $D=4$ and physical dimension $D=2$, which can be seen from
Tables~4 and 5.

In a few  exceptional cases either  $\tilde{\gamma}_1$ or $\tilde{\gamma}_2$ are zero.
In these cases interaction is absent or reduced to splitting of the pomeron
 into two odderons. With such interaction the full propagators can be found exactly and do not
possess any singularity at $\delta_{1,2}$, which can be seen directly. With such behaviour the masses can be continued to negative values
without difficulty. However, this leads to  renormalized propagators growing with energy,
which prohibits the perturbative treatment  and prohibits  our approach. Then it is not the renormalization group that
is to be applied but rather summation of multiple reggeon exchanges should be attempted, as in the very old approach
of A.D.Kaidalov and K.A.Ter-Martirosyan ~\cite{kater}.

In Section 6 we calculated the asymptotical behaviour of the elastic scattering amplitudes at high energies $\sqrt{s}$.
We adopted the same assumption that was made in ~\cite{abarb2} without odderons, namely that the coupling of the
participant hadrons to the pomeron-odderon system do not depend on transferred energies and so has a quasi Glauber structure.
The found asymptotical amplitude is described by  single exchange of either the full pomeron Green function or the odderon one.
The dominant pomeron part is found the same as in ~\cite{abarb2} leading to the cross-section growing as
$(\ln s)^{1/6}$. Odderon do not change this leading behaviour in the
positive signature amplitude. The  odderon part with the negative signature is found subdominant but also leading to the
rising cross-section as $(\ln s)^{1/12}$.

Comparing the found cross-sections with the existing experimental data,
one has to take into account two essential points similar to~\cite{abarb2}.
First, our predictions refer to asymptotically high energies, perhaps, close
to the Froissart limit, which is still far away from the attained experimentally.
At lower energies the cross-sections derived in our model can be very different
from their found asymptotic form. Second and more important, one has to take
into account approximations done in the course of derivation. Intrinsic to the
renormalization group approach is that the model is critical in $D=4$ dimensions,
while in reality it lives in $D=2$. So all our results are initially obtained
at $D=4-\epsilon$ and then continued to $D=2$. To make this continuation more
reliable one has to find results in the form of a series in powers of
$\epsilon$ and then try to sum it, probably, using something like
the Borel summation. However, this requires going beyond the single loop
approximation, what lies outside the scope of the present article.

One may also ask how our results are related to the QCD picture. Actually,
the QCD is oriented to the so-called ''hard'' processes with small interparton
distances. Total cross-sections lie outside its scope. Attempts to study them,
say, in the well-known BFKL approach give cross-sections rising as a power
of energy and so certainly wrong at high energies. Our treatment also gives
rising cross-sections but compatible with the Froissart restriction. So they
are, at least, satisfactory in a qualitative manner.



\newpage
\section{Appendix 1. Real fixed points}
In ~\cite{bkv2023}   we found five real fixed points,
four with $g_3=0$: $g_c^{(0)}$  $g_c^{(1)}$, $g_c^{(2)}$, $g_c^{(3)}$
and one $g_c^{(4)}$ with $g_1=g_2=0$. Due to singularity of the $\beta$-functions
the fixed points at which  $g_4=0$ also have $g_3=0$. Then they are characterized
by the ratio $r=g_3/g_4$, which may also be zero or have a fixed finite value.
The fixed points are  presented in Table~1 in which we show the
corresponding coupling constants at the fixed point with $g_{1,2,3}$
divided by $\sqrt{\ep}$.

\begin{center}
{\bf Table 1.}\\
\vspace{0.2cm}
Coupling constants at fixed points divided by $\sqrt{\ep}$\\
\vspace{0.2cm}
\begin{tabular}{|r|r|r|r|r|}
\hline
fixed point&$g_1$&$g_2$&$g_3$&$g_4$\\\hline
$g_c^{(0)}$&0&0&$0$&0\\\hline
$g_c^{(1)}$&$\frac{1}{\sqrt{6}}$&0&0&0\\\hline
$g_c^{(2)}$&$\frac{1}{\sqrt{6}}$&0.39750&0&0.88961\\\hline
$g_c^{(3)}$&$\frac{1}{\sqrt{6}}$&$\frac{1}{\sqrt{96}}$&0&0\\\hline
$g_c^{(4)}$&0&0&2&2\\
\hline
\end{tabular}
\end{center}
At $g_c^{(0)}$ ratio $r=g_3/g_4=1$, at $g_c^{(1)}$ and at $g_c^{(3)}$ $r=0$.

Attraction or repulsion at the fixed points is described by the matrix $a_{ij}=\partial \beta_i/\partial g_j$, $i,j=1,...,4$.
Its eigenvectors for positive eigenvalues indicate direction along which the fixed point is attractive, those for
negative eigenvalues show directions along with the fixed point is repulsive. The number of positive and negative eigenvalues
is different for different fixed points. In Table~2 we show  eigenvalues $x=\{x_1,x_2, x_3,x_4\}$ for  matrix $2a$ at $\ep=2$.
Zero eigenvalues describe directions along which the corresponding projection of the 4-vector $g_i$ does not move
in the vicinity of the fixed point and stays equal to its initial value.
\begin{center}
{\bf Table 2.}\\
\vspace{0.2cm}
Eigenvalues of  matrix $2a$ at $\ep=2$\\
\vspace{0.2cm}
\begin{tabular}{|r|r|r|r|r|}
\hline
fixed point&$x_1$&$x_2$&$x_3$&$x_4$\\\hline
$g_c^{(0)}$&0&-2&-2&0\\\hline
$g_c^{(1)}$&2&-1&-1&1/6\\\hline
$g_c^{(2)}$&2&1.2085&0.36956&-0.13976\\\hline
$g_c^{(3)}$&2&1&1/6&1/(6u)\\\hline
$g_c^{(4)}$&2&-16/3&2&2\\
\hline
\end{tabular}
\end{center}
Note that of all fixed points only $g_c^{(3)}$ is purely attractive. All the rest have one or several repelling
directions, so that arriving at them is only possible in a restricted domain of all coupling constants.
To find the probability of arriving at concrete fixed points in ~\cite{bkv2023} we studied the trajectories starting from some points
(outside the fixed ones) distributed homogeneously  in a hypercube of the four coupling constants around zero.
We studied 185 000 trajectories. Since we were working in the single loop approximation in 35 \% of all cases it was impossible to follow
the trajectories far away from the fixed point and so they went to infinitely large values of coupling constants
("to infinity'). In the rest 65 \% cases the distribution  of the arrival at specific fixed poits was found to be in percentage
\[g_c^{(0)}:g_c^{(1)}:g_c^{(2)}:g_c^{(3)}:g_c^{(4)} \ =\ 0:0.33:0:92.6:7.1\ .\]
As expected in the vast majority of cases the trajectories arrive at the purely attractive fixed point $g_c^{(3)}$.

Next we present coefficients $d_i$ for each fixed point.

\begin{center}
{\bf Table 3.}\\
\vspace{0.2cm}
Coefficients $d_i$ at fixed points\\
\vspace{0.2cm}
\begin{tabular}{|r|r|r|r|}
\hline
fixed point&$d_1$&$d_2$&$d_3$\\\hline
$g_c^{(0)}$&0&0&$-\frac{1}{2}$\\\hline
$g_c^{(1)}$&$\frac{1}{12}$&0&0\\\hline
$g_c^{(2)}$&$\frac{1}{12}$&0.17701&0\\\hline
$g_c^{(3)}$&$\frac{1}{12}$&$\frac{1}{24}$&0\\\hline
$g_c^{(4)}$&0&0&$-\frac{1}{2}$\\
\hline
\end{tabular}
\end{center}

Finally, we show
the anomalous dimensions  necessary for
the construction of $\Gamma^R_i$ at small $\delta_1$, $\delta_2$ or $E$.


\begin{center}
{\bf Table 4.}\\
\vspace{0.2cm}
Anomalous dimensions
$\tilde{\gamma}_{1,2}$, $\zeta-1$ and $z-1$  at fixed points for small $\delta_1$\\
\vspace{0.2cm}
\begin{tabular}{|r|r|r|r|r|}
\hline
fixed point&$\tilde{\gamma}_1$&$\tilde{\gamma}_2$&$\zeta-1$&$z-1$\\\hline
$g_c^{(0)}$&0&$-\frac{\ep/2}{1-\ep/2}$&$\frac{\ep/2}{1-\ep/2}$&$0$\\\hline
$g_c^{(1)}$&$-\frac{\ep/6}{1-\ep/12}$&$-\frac{\ep/12}{1-\ep/12}$&$\frac{\ep/12}{1-\ep/12}$&$\frac{\ep/8}{1-\ep/12}$\\\hline
$g_c^{(2)}$&$-\frac{\ep/6}{1-\ep/12}$&$-\frac{0.26034\ep}{1-\ep/12}$&$\frac{\ep/12}{1-\ep/12}$&$\frac{\ep/8}{1-\ep/12}$\\\hline
$g_c^{(3)}$&$\frac{-\ep/6}{1-\ep/12}$&$-\frac{\ep/8}{1-\ep/12}$&$\frac{\ep/12}{1-\ep/12}$&$\frac{\ep/8}{1-\ep/12}$\\\hline
$g_c^{(4)}$&0&$-\frac{\ep/2}{1-\ep/2}$&$\frac{\ep/2}{1-\ep/2}$&$\frac{\ep/2}{1-\ep/2}$\\
\hline
\end{tabular}
\end{center}


\begin{center}
{\bf Table 5.}\\
\vspace{0.2cm}
Anomalous dimensions
$\gamma_{1,2}$, $\zeta-1$ and $z-1$, divided by $\ep$ at fixed points for $\delta_2\to 0$\\
\vspace{0.2cm}
\begin{tabular}{|r|r|r|r|r|}
\hline
fixed point&$\gamma_1$&$\gamma_2$&$\zeta-1$&$z-1$\\\hline
$g_c^{(0)}$&$1/2$&0&-1/2&$-1/2$\\\hline
$g_c^{(1)}$&-1/12&0&-1/12&1/24\\\hline
$g_c^{(2)}$&-1/12&-0.17701&-1/12&1/24\\\hline
$g_c^{(3)}$&-1/12&-1/24&-1/12&1/24\\\hline
$g_c^{(4)}$&1/2&0&-1/2&0\\
\hline
\end{tabular}
\end{center}

\begin{center}
{\bf Table 6.}\\
\vspace{0.2cm}
Anomalous dimensions
$\gamma_{1,2}$, $\tau_1$ divided by $\ep$ and $z$ at $D=2$ at fixed points for $E\to 0$\\
\vspace{0.2cm}
\begin{tabular}{|r|r|r|r|r|}
\hline
fixed point&$\gamma_1$&$\gamma_2$&$\tau_1$&$z$ at $D=2$\\\hline
$g_c^{(0)}$&1/2&0&1/2&0\\\hline
$g_c^{(1)}$&-1/12&0&-1/24&13/12\\\hline
$g_c^{(2)}$&-1/12&-1/11.3&-1/24&13/12\\\hline
$g_c^{(3)}$&-1/12&-1/24&-1/24&13/12\\\hline
$g_c^{(4)}$&1/2&0&0&1\\
\hline
\end{tabular}
\end{center}

\section{Appendix 2. Calculation of scaling functions}
\subsection{$\delta_1\to 0$}
We first consider the case of small $\delta_1$.

At $\ep=0$ we have
$\zeta(0)=z(0)=1$, so at zero order
\beq
\rho_{1,\ep=0}=\rho_{10}=\frac{E}{\delta_1},\ \ \rho_{2,\ep=0}=\rho_{20}=\frac{\alpha'_1 k^2}{\delta_1},\ \
\rho_{3,\ep=0}=\rho_{30}=\frac{\delta_2}{\delta_1}.
\label{rho0}
\eeq

Using (\ref{phie})
in  the first two orders in $\ep$   for both pomeron and odderon we can present
\[
\Gamma_j^R(E,k^2,g_c(\ep),\alpha'_1,\delta_1,\delta_2,E_N)=
\delta_1\Big(\frac{\delta_1}{E_N}\Big)^{-\tilde{\gamma}_j}\Big\{\Phi_0(\rho_{i0})\]\[+
\ep\Big[\Phi_1(\rho_{i0})-\tilde{\gamma}'(0)L\Phi_0(\rho_{i0})
-\zeta'(0)L\rho_{10}\frac{\pd\Phi(\rho_{i0})}{\pd\rho_{10}}\]
\beq
-z'(0)L\rho_{20}\frac{d\Phi(\rho_{i0})}{\pd\rho_{20}}
-\zeta'(0)L\rho_{30}\frac{\pd\Phi(\rho_{i0})}{\pd\rho_{30}}\Big]\Big\}.
\label{scfun}
\eeq
Here we used the notation
\[ L=t=\ln\frac{\delta_1}{E_N}.\]
Our strategy will be to find $\Phi_0$ and $\Phi_1$ from this expression comparing it with
the direct development of the Green functions in powers of $\ep$ up to linear terms.

We start from the zeroth order $\ep=0$. The inverse propagators are
for the pomeron
\[\Gamma^R_1=E-\alpha'_1k^2-\delta_1\]
and for the odderon
\[\Gamma^R_2=E-\alpha'_2k^2-\delta_2.\]
Since $E=\rho_{10}\delta_1$,  $\alpha'_1k^2=\rho_2\delta_1$ and $\delta_2=\delta_1\rho_{30}$,
we find
\[\Gamma^R_1=\delta_1(\rho_{10}-\rho_{20}-1)\]
and
\[\Gamma^R_2=\delta_1(\rho_{10}-u\rho_{20}-\rho_{30}).\]
So in the lowest order we get (\ref{phi10}) and (\ref{phi20}).

In the linear order in $\ep$
\[\Gamma^R_j(E,k^2,g_i(\epsilon),\alpha'_1,E_N)_{linear\ in\ \ep}=
-\Sigma^R_j, \ \ j=1,2.\]
The renormalized self-mass for the pomeron is
$\Sigma^R_1=\Sigma^R_a+\Sigma^R_b$,
with
\[\Sigma^R_a=\epsilon d_1\Gamma(2-D/2)
\sigma_1\Big(\frac{(\sigma_1/E_N)^{D/2-2}}{1-D/2}+1\Big),\]
\[
\Sigma^R_b=\epsilon d_3\Gamma(2-D/2)\Big[
\sigma_3\Big(\frac{(\sigma_3/E_N)^{D/2-2}}{1-D/2}+1\Big)-
2\delta_2\Big(\frac{(2\delta_2/E_N)^{D/2-2}}{1-D/2}+1\Big)\Big],\]
For the odderon the renormalized self-mass is
\[
\Sigma^R_2=\epsilon d_2\Gamma(2-D/2)
\Big[\sigma_2\Big(\frac{(\sigma_2/E_N)^{D/2-2}}{1-D/2}+1\Big)
-\delta_1\Big(\frac{(\delta_1/E_N)^{D/2-2}}{1-D/2}+1\Big)\Big].\]
Here
$\sigma_i$, $i=1,2,3$ are given by Eqs.~(\ref{sms1}), (\ref{sms2}) and (\ref{sms3}).
The constants $d_i$ are defined by (\ref{defd123}).
At $\ep\to 0$ these expressions simplify.
We use
\[\Gamma(2-D/2)=\frac{2}{\ep},\ \ \frac{a^{D/2-2}}{1-D/2}=-1+\frac{\ep}{2}(\ln a-1)\]
to get
\beq
\Gamma(2-D/2)\frac{a^{D/2-2}}{1-D/2}+1=\ln a+1.
\label{eto0}
\eeq

Using (\ref{eto0}) we find in the limit $\ep\to 0$
\[\Sigma^R_a=\epsilon d_1\sigma_1
\Big(\ln\frac{\sigma_1}{E_N}-1\Big),\]
\[
\Sigma^R_b=\epsilon d_3\Big[
\sigma_3\Big(\ln\frac{\sigma_3}{E_N}-1\Big)-
2\delta_2\Big(\ln\frac{2\delta_2}{E_N}-1\Big)\Big],\]
\[
\Sigma^R_2=\epsilon d_2
\Big[\sigma_2\Big(\ln\frac{\sigma_2}{E_N}-1\Big)
-\delta_1\Big(\ln\frac{\delta_1}{E_N}-1\Big)\Big].\]

We express $\sigma_i$ via $\rho_{i0}$ defining $\sigma_i=\delta_1 x_i$ with
$x_i$ given by (\ref{defxi1}) at zero order in $\ep$, that is via $\rho_{i0}$,
and rewrite the self-mass for the pomeron as
\[\Sigma^R_a=\epsilon \delta_1d_1x_1(L+\ln x_1-1),\]
\[
\Sigma^R_b=\epsilon \delta_1d_3
\Big[x_3(L+\ln x_3-1)-2\rho_{30}(L+\ln (2\rho_{30})-1)\Big],\]
so that
\[\Sigma_1^R=\ep\delta_1\Big(d_1x_1(\ln x_1-1)+d_3x_3(\ln x_3-1)-2d_3\rc(\ln(2\rc)-1)\Big)+\ep\delta_1Y_1,\]
where
\[ Y_1=L(d_1x_1+d_3x_3-2d_3\rc)=L\Big[d_1\Big(\frac{1}{2}\rb+2-\ra\Big)+d_3\Big(\frac{1}{2}u\rb-\ra\Big]\Big]\]\beq
=L\Big[-\ra\Big(d_1+d_3\Big))+\rb\Big(\frac{1}{2}d_1+\frac{u}{2}d_3\Big)+2d_1\Big].
\label{y1}
\eeq

For the odderon we get
\[\Sigma^R_2=\epsilon \delta_1d_2\Big[x_2(\ln x_2-1)+1)\Big]+\ep\delta_1Y_2,\]
where
\beq
Y_2=d_2L\Big(\frac{u}{1+u}\rb+\rc-\ra\Big).
\label{y2}
\eeq

In the linear order in $\ep$ we should have
\[
-\Sigma^R_i=\ep\delta_1\Phi_{i1}-\ep\delta_1 X_i,\ \ i=1,2,\]
where
\beq
X_i=\tilde{\gamma}'_i(0)L\Phi_{i0}(\rho_{i0})
+\zeta'(0)L\rho_{10}\frac{d\Phi_{i0}(\rho_{i0})}{\pd\rho_{10}}
+z'(0)L\rho_{20}\frac{d\Phi_{i0}(\rho_{i0})}{\pd\rho_{20}}
+\zeta'(0)L\rho_{30}\frac{d\Phi_{i0}(\rho_{i0})}{\pd\rho_{30}}.
\label{xxi}
\eeq
So the scaling function linear in $\epsilon$ is given by
\beq\ep\delta_1\Phi_{i1}=\ep\delta_1X_i-\Sigma^R_i.
\label{phii1}
\eeq

The coefficients in (\ref{xxi}) up to terms linear in $\ep$ are obtained as follows.
\[\gamma_1=-\ep(d_1+d_3),\ \ \gamma_2=-\ep d_2,\ \ \kappa_1=\ep(d_1-d_3),\]
\[\tilde{\gamma}_1=-2\ep d_1,\ \ \tilde{\gamma}_2=\ep(-d_1-d_2+d_3),\]
\[\tau_1=-\ep\frac{1}{2}d_1+\ep\frac{1}{2}(2-u)d_3,\ \ z=1+\ep\Big(\frac{3}{2}d_1-\frac{u}{2}d_3\Big),\]
\[\zeta=1+\kappa_1=1+\ep(d_1-d_3).\]

We start with $\Phi_{11}$.
We get
\[X_1=L\Big[-2d_1\Big(\ra-\rb-1\Big)+\Big(d_1-d_3\Big)\ra-\Big(\frac{3}{2}d_1-\frac{u}{2}d_3\Big)\rb\Big]\]\[
=L\Big[-\ra\Big(d_1+d_2\Big)+\rb\Big(\frac{1}{2}d_1+\frac{u}{2}\Big)d_3+2d_1\Big].\]
One observes that
$Y_1-X_1=0$,
so that we find (\ref{phi11}).

Now we consider $\Phi_{21}$.
We have
\[X_2=L\Big[(-d_1-d_2+d_3)(\ra-u\rb-\rc)+(d_1-d_2)\ra -u\Big(\frac{3}{2}d_1-\frac{u}{2}d_3)\Big)\rb-(d_1-d_3)\rc\Big]\]\[
=L\Big[ -d_2\ra+\rb\Big(-\frac{1}{2}ud_1+ud_2+d_3-u+u^2/2\Big)+\rc d_2\Big].\]
So we find
\beq
X_2-Y_2=L\rc\Big(-\frac{1}{2} ud_1+d_2\frac{u^2}{1+u}+d_3u(u/2-1)\Big).\label{xmy}\eeq
Multiplied by $\ep$ the bracket is
\[  -\frac{1}{4}g_1^2+\frac{u^2g_2^2}{(1+u)[(1+u)/2]^{D/2}}-\frac{u(u-2)g_3^2}{4u^{D/2}}\]
and at $D=4$ (or $\ep=0$) is
\[-\frac{u}{4}g_1^2+\frac{4u^2g_2^2}{(1+u)^3}-\frac{(u-2)g_3^2}{4u}=-\beta_4.\]
So at the fixed point it is equal to zero and $X_2-Y_2=0$.
As a result, we get (\ref{phi21}).

Note that cancelling of terms containing $L=\ln\delta_1/E_N$ follows from scaling, which prohibits extra
arguments in $\Phi$ apart from $\rho_i$, $i=1,2,3$.

This ends calculations of $\Phi$ for small $\delta_1$.

\subsection{$\delta_2\to 0$}
Passing to the construction of the scaling functions as a
series in small $\ep$ we
introduce as before
\[\rho_{10}=\rho_{1}=\frac{E}{\delta_2} ,\ \ \rho_{20}=\frac{\alpha'_1k^2}{\delta_2},\ \ \rho_{30}=\frac{\delta_1}{\delta_2}.\]
Since now evolution of $E$ does  not depend on $\ep$, Eq.~(\ref{scfun}) somewhat simplifies to
\[
\Gamma^R(E,k^2,g_c(\ep),\alpha'_1,\delta_1,\delta_2,E_N)=
\delta_2\Big\{\Phi_0(\rho_{i0})\]\beq+
\ep\Big[\Phi_1(\rho_{i0})-{\gamma}'(0)L\Phi_0(\rho_{i0})
-z'(0)L\rho_{20}\frac{\pd\Phi(\rho_{i0})}{\pd\rho_{20}}-\zeta'(0)L\rho_{30}\frac{\pd\Phi(\rho_{i0})}{\pd\rho_{30}}\Big]\Big\},
\label{scfuna}
\eeq
where now
\[L= t=\ln\frac{\delta_2}{E_N}.\]
We find at $\ep=0$
\[\Gamma^R_1=\delta_2(\rho_{10}-\rho_{20}-\rho_{30}),\ \
\Gamma^R_2=\delta_2(\ra-u\rb-1),\]
which allows to derive (\ref{phi10a}) and (\ref{phi20a}).

Next we pass to terms linear in $\ep$. Now we separate from $\sigma_i$
defined by (\ref{sms1}), (\ref{sms2}) and (\ref{sms3}) factor $\delta_2$
putting $\sigma_i=\delta_2 x_i$, $i=1,2,3$, where now they are defined in  (\ref{defxi2})
and rewrite the self-mass for the pomeron as
\[\Sigma^R_a=\epsilon \delta_2d_1x_1(L+\ln x_1-1),\]
\[
\Sigma^R_b=\epsilon \delta_2d_3
\Big[x_3(L+\ln x_3-1)-2(L+\ln 2-1)\Big],\]
so that
\[\Sigma_1^R=\ep\delta_2\Big(d_1x_1(\ln x_1-1)+d_3x_3(\ln x_3-1)
-2d_3(\ln 2-1)\Big)+\ep\delta_2Y_1,\]
where
\[ Y_1=L\Big(d_1x_1+d_3(x_3-2)\Big)=
L\Big[d_1\Big(\frac{1}{2}\rb+2\rc-\ra\Big)
+d_3\Big(\frac{1}{2}u\rb-\ra\Big)\Big]\]\beq
=L\Big[-\ra\Big(d_1+d_3\Big)+\rb\Big(\frac{1}{2}d_1+\frac{u}{2}d_3\Big)
+2d_1\rc\Big].
\label{y1a}
\eeq

For the odderon we get
\[\Sigma^R_2=\epsilon \delta_2d_2\Big[x_2(\ln x_2-1)
-\rc(\ln\rc-1)\Big]
+\ep\delta_2Y_2,\]
where
\beq
Y_2=d_2L\Big(\frac{u}{1+u}\rb+1-\ra\Big).
\label{y2a
}
\eeq

In order $\ep$
\[
\ep\delta_2\Phi_{i1}=\ep\delta_2X_i-\Sigma^R_i,\]
where
\beq
X_i=
{\gamma}'_i(0)L\Phi_{i0}(\rho_{i0})
+z'(0)L\rho_{20}\frac{\pd\Phi_{i0}(\rho_{i0})}{\pd\rho_{20}}
+\zeta'(0)L\rho_{30}\frac{\pd\Phi_{i0}(\rho_{i0})}{\pd\rho_{30}}.
\eeq
Here
\[\frac{\pd\Phi_{10}(\rho_{i0})}{\pd\rho_{20}}=-1,\ \
\frac{\pd\Phi_{20}(\rho_{i0})}{\pd\rho_{20}}=-u,\ \
\frac{\pd\Phi_{10}(\rho_{i0})}{\pd\rho_{30}}=-1,\ \
\frac{\pd\Phi_{20}(\rho_{i0})}{\pd\rho_{30}}=0.\]
The coefficients are
\[
{\gamma}'_1=-d_1-d_3,\ \ {\gamma}'_2=-d_2,\]
\[z'(0)=\frac{1}{2}d_1+\frac{1}{2}(2-u)d_3,\ \ \zeta'(0)=d_3-d_1 .\]
So we get
\[X_1=L\Big[(-d_1-d_3)(\ra-\rb-\rc)-
\Big(\frac{1}{2}d_1+\frac{1}{2}(2-u)d_3\Big)\rb-\rc(d_3-d_1)\Big]\]\[=
L\Big[\ra\Big(-d_1-d_3\Big)+\rb\Big(\frac{1}{2}d_1+\frac{u}{2}d_3\Big)
+2d_1\rc\Big)\Big].\]
\[X_2=L\Big[-d_2(\ra-u\rb-1)-
u\rb\Big(\frac{1}{2}d_1+\frac{1}{2}(2-u)d_3\Big)\Big]\]\[
=L\Big[-d_2\ra+\rb\Big(-\frac{u}{2}d_1+ud_2-\frac{u}{2}(2-u)d_3\Big)
+d_2\Big].\]

As a result, we find that both differences are zero, as in the previous case  case when one scales $\delta_1$:
$X_i-Y_i=0,\ \ i=1,2$. So in the end we get (\ref{phi11a}) and (\ref{phi21a}).

\section{Appendix 3. Disconnected pieces in the Green function}
Consider the case when the Green function splits into two disconnected parts $G_1$ and $G_2$. For given number of participants it is shown in
Fig.~3.
\begin{figure}
\begin{center}
\epsfig{file=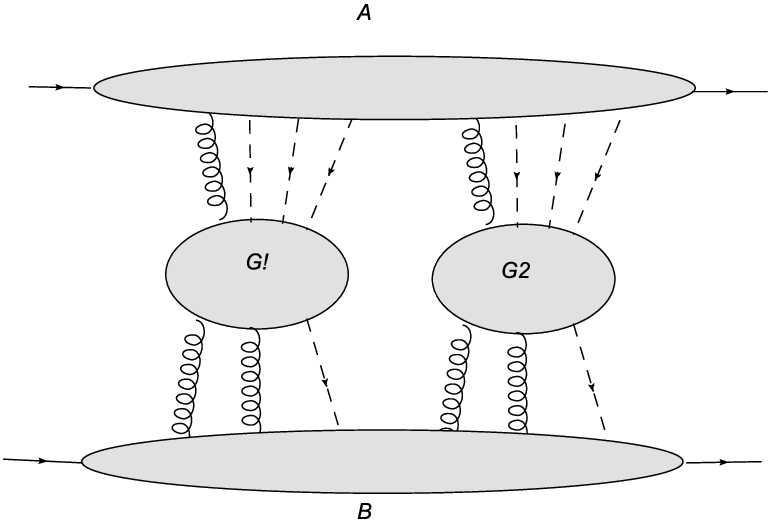, width=8 cm}
\caption{Elastic amplitude when the Green function contains two disconnected parts.}
\end{center}
\label{fig3}
\end{figure}
We shall denote variables pertaining to $G_1$ and $G_2$ by upper indices $(1)$ and $(2)$. So the initial and final numbers of reggeons
for the two connected parts will be $n^{(1)}$, $n^{(2)}$ and    $m^{(1)}$, $m^{(2)}$' The total number of the initial reggeons
will be $n=n^{(1)}+n^{(2)}$ and  $m=m^{(1)}+m^{(2)}$. Similarly
numbers of pomerons and
odderons will be $n_P^{(1)}$, $n_P^{(2)}$ and  $n_O^{(1)}$, $n_O^{(2)}$ and their total number in the whole diagram will be
$n_P=n_P^{(1)}+n_P^{(2)}$ and $n_O=n_O^{(1)}+n_O^{(2)}$. The overall total number of reggeons in the whole diagram will
evidently be $n_t=n_t^{(1)}+n_t^{(2)}=n+m=n_P+n_O$.

The contribution with given numbers of reggeons will be written as
\[
{\cal A}^{(nm)}=A^{(n)}B^{(m)}\int d\tau_1d\tau_2dE^{(1)}dE^{(2)}d^Dq^{(1)}d^Dq^{(2)}
\delta (E^{(1)}+E^{(2)}-E)\delta^D(q^{(1)}+q^{(2)}-q)G^{(nm)},\]
where now index
\[(nm)=n_1^{(1)}n_2^{(1)}n_1^{(2)}n_2^{(2)}
m_1^{(1)}m_2^{(1)}m_1^{(2)}m_2^{(2)}\]
includes all numbers of initial and final
reggeons. Each of the phase volumes $d\tau_1$ and $d\tau_2$ is the same as in (\ref{eq0110}) with variables belonging to  each of
the two connected parts of the Green function.

For each of the two disconnected part of $G^{(nm)}$ we can write the scaling property (\ref{eqad2})
\[
G_R^{(nm)_1}(E_i,k_i)=E_N^{1-n_t^{(1)}}\Big(\frac{E_N}{\alpha'_1}\Big)^{(2-n_t^{(1)})D/4}\xi_1^{c_1}
\Phi^{(nm)_1}\Big(-\frac{E_i}{E^{(1)}},\xi_1^{-z}\frac{{k}_i{k}_j}{E_N}\alpha'_1,g_c\Big)\]
\beq
\prod_{i=1}^{n_P^{(1)}}\Big[\Phi_{1}\Big(\xi_1^{-z}\frac{k_i^2}{E_N},g_c\Big)\Big]^{-1}
\prod_{i=1}^{n_O^{(1)}}\Big[\Phi_{2}\Big(\xi_1^{-z}\frac{k_i^2}{E_N},g_c\Big)\Big]^{-1}
\label{eqad21}\eeq
and
\[
G_R^{(nm)_2}(E_i,k_i)=E_N^{1-n_t^{(2)}}\Big(\frac{E_N}{\alpha'_1}\Big)^{(2-n_t^{(2)})D/4}\xi_2^{c_2}
\Phi^{(nm)_2}\Big(-\frac{E_i}{E^{(2)}},\xi_2^{-z}\frac{{k}_i{k}_j}{E_N}\alpha'_1,g_c\Big)\]
\beq
\prod_{i=1}^{n_P^{(2)}}\Big[\Phi_{1}\Big(\xi_2^{-z}\frac{k_i^2}{E_N},g_c\Big)\Big]^{-1}
\prod_{i=1}^{n_O^{(2)}}\Big[\Phi_{2}\Big(\xi_2^{-z}\frac{k_i^2}{E_N},g_c\Big)\Big]^{-1}.
\label{eqad22}\eeq
Here
\[\xi_1=\frac{-E^{(1)}}{E_N},\ \ \xi_2=\frac{-E^{(2)}}{E_N},\]
\[c_1=1-n_t^{(1)}+\frac{1}{2}\gamma_1 n_P^{(1)}+\frac{1}{2}\gamma_2n_O^{(1)}+z(2-n_t^{(1)})\frac{D}{4},\]
\[c_2=1-n_t^{(2)}+\frac{1}{2}\gamma_1 n_P^{(2)}+\frac{1}{2}\gamma_2n_O^{(2)}+z(2-n_t^{(2)})\frac{D}{4}\]
and variables $E_i$ and ${k}_i$  belong to $G_1$ in (\ref{eqad21}) and to $G_2$ in (\ref{eqad22}).

As before,  we make a change of integration variables
\[E_i^{(1)}=E\zeta_i^{(1)}.\ \ E_i^{(2)}=E\zeta_i^{(2)},\ \ E^{(1)}=E\zeta^{(1)},\ \ E^{(2)}=E\zeta^{(2)}.\]
The total number of integrations is $n_t+2$. However, we have five $\delta$-functions. So from this change we have factor
$E^{n_t-3}$
and integrations over $\zeta$ will be constrained to have $\sum\zeta_i^{(1,2)}=\zeta^{(1,2)}$
and $\zeta^{(1)}+\zeta^{(2)}=1$.

Next we change
\[{k}^{(1)}_i=\xi^{z/2}{x}_i^{(1)},\ \ {k}^{(2)}_i=\xi^{z/2}{x}_i^{(2)},\
{q}^{(1)}=\xi^{z/2}{x}^{(1)},\ \ {q}^{(2)}=\xi^{z/2}{x}^{(2)}.\]
Taking into account the relevant five $\delta$-functions we obtain factor
$ \xi^{D(n_t-3)z/2}$.
The relevant  constraint on ${x}$ are
\beq
\sum_i{x}_i^{(1,2)}={x}^{(1,2)},\ \ {x}^{(1)}+{x}^{(2)}={q}\xi^{-z/2}.
\label{eqad9}\eeq

Turning to our Green functions $G_1$ and $G_2$ we find in (\ref{eqad21})
$E_i/E^{(1)}=\zeta^{(1)}_i/\zeta^{(1)}$.
Further,
\[\xi_1^{-z/2}{k}_i^{(1)}=\Big(\frac{\xi}{\xi_1}\Big)^{-z/2}{x}^{(1)}_i={\zeta^{(1)}}^{z/2}{x}^{(1)}_i,\]
so that functions $\Phi$ depend only on our new variables $\zeta$ and ${x}$.
The same is true for $G_2$ in (\ref{eqad22}).

In the product $G_1G_2$ the dependence on $E$ and ${q}$ becomes concentrated in the factor
\[\xi_1^{c_1}\xi_2^{c_2}=E^{c_1+c_2}\Big(\frac{-\zeta^{(1)}}{E_N}\Big)^{c_1}\Big(\frac{-\zeta^{(2)}}{E_N}\Big)^{c_2}.\]
Separating the $E$-factor we find that the product $G_1G_2$ can be presented as
\[G_1G_2=E^{c_1+c_2}Q(\zeta,{x})\]
with some function $Q$ which depends only on our new variables $\zeta$ and ${x}$.
Integration over these new variables wiill add factor
$E^{n_t-3}\xi^{D(n_t-3)z/2}$
and the result of this integration will only depend on $q^2\xi^{-z}$ due to the delta function (\ref{eqad9}).

So we finally find
\[
{I}^{(nm)}=E^{-1+a}F^{(nm)}(t\xi^{-z}),\]
where
\[-1+a=n_t-3+D(n_t-3){z/2}+c_1+c_2\]\beq
=-1+\frac{1}{2}\gamma_1n_P+\frac{1}{2}\gamma_2n_O+\frac{1}{4}zD(n_t-2).\label{eqad8}\eeq
This is the same expression (\ref{eqad4}) as would be obtained if the Green function was connected with the
same $n_t$, $n_P$ and $n_O$. So division of the Green function into disconnected parts does not influence our results.


\end{document}